\documentclass[manuscript]{acmart}

\AtBeginDocument{%
  \providecommand\BibTeX{{%
    \normalfont B\kern-0.5em{\scshape i\kern-0.25em b}\kern-0.8em\TeX}}}

\setcopyright{acmcopyright}
\copyrightyear{2023}
\acmYear{2023}
\acmDOI{10.1145/1122445.1122456}

\acmPrice{15.00}
\acmISBN{978-1-4503-XXXX-X/18/06}

\newcommand{\re}[1]{{\color{black} #1}}

\newcommand\conIC{\texttt{\textcolor{black}{In-Context}}}
\newcommand\conPar{\texttt{\textcolor{black}{List}}}
\newcommand\conSeq{\texttt{\textcolor{black}{Dropdown}}}

\usepackage{multirow}
\usepackage{subcaption}
\usepackage{caption}
\usepackage{tablefootnote}

\usepackage{algorithm}
\usepackage{algorithmicx}
\usepackage{algpseudocode}

\begin{document}

% \title[When Diverse Initial Examples Lead to Better Design Solutions?]{When Diverse Initial Examples Lead to Better Design Solutions?}
% \title[Exploring Presentation Interaction Effects on Mechanism of Inspiration from Diverse Examples]{Exploring Presentation Interaction Effects on Mechanism of Inspiration from Diverse Examples}
\title[Interface Presentation Effects on Example-Based Problem Solving]{Formulating or Fixating: Effects of Examples on Problem Solving Vary as a Function of Example Presentation Interface Design}

\author{Joel Chan}
\affiliation{%
  \institution{College of Information Studies, University of Maryland}
  \country{USA}}
  
\author{Zijian Ding}
\affiliation{%
  \institution{\institution{College of Information Studies, University of Maryland}
  \country{USA}}}
  
\author{Eesh Kamrah}
\affiliation{%
  \institution{Department of Mechanical Engineering, University of Maryland}
  \country{USA}}
  
\author{Mark Fuge}
\affiliation{%
  \institution{Department of Mechanical Engineering, University of Maryland}
  \country{USA}}
  
% \author{Anonymous}
% \affiliation{%
%   \institution{Anonymous}}
  
% \author{Anonymous}
% \affiliation{%
%   \institution{Anonymous}}

\renewcommand{\shortauthors}{Chan, et al.}

\begin{abstract}
Interactive systems that facilitate exposure to examples can augment problem solving performance. However designers of such systems are often faced with many practical design decisions about how users will interact with examples, with little clear theoretical guidance. To understand how example interaction design choices affect whether/how people benefit from examples, we conducted an experiment where 182 participants worked on a controlled analog to an exploratory creativity task, with access to examples of varying diversity and presentation interfaces. Task performance was worse when examples were presented in a list, compared to contextualized in the exploration space or shown in a dropdown list. Example lists were associated with more fixation, whereas contextualized examples were associated with using examples to formulate a model of the problem space to guide exploration. We discuss implications of these results for a theoretical framework that maps design choices to fundamental psychological mechanisms of creative inspiration from examples.
\end{abstract}

 
%  </ccs2012>

% \ccsdesc[500]{Human-centered computing~Empirical studies in HCI}
% \ccsdesc[500]{Human-centered computing~Auditory feedback}
% \ccsdesc[300]{Human-centered computing~Empirical studies in interaction design}
% \ccsdesc[500]{Human-centered computing~Sound-based input / output}


\maketitle

\def \RQA {With what presentations do initial examples lead to better final creative problem solutions?}

\def \RQB {How do different initial examples influence creator’ behaviors of exploring creative problem solutions?}

\def \stimulate {stimulation-based}

\def \model {model-based}

\section{Introduction}
\label{sec:intro}

Examples --- \re{descriptions or representations of possible \textit{solutions} (or parts thereof) for the same or related problems \cite{sioFixationInspirationMetaanalytic2015,crillyWhereNextResearch2017,herringGettingInspiredUnderstanding2009,smithConstrainingEffectsExamples1993,brandtExamplecentricProgrammingIntegrating2010}} --- are an integral part of the creative problem solving process. \re{Examples can take many forms, such as previous physical prototypes brought to a brainstorm \cite{hargadonTechnologyBrokeringInnovation1997}, search results for patents for related problems \cite{fuMeaningFarImpact2013}, spoken ideas from collaborators \cite{nijstadHowGroupAffects2006}, UI designs in an accessible gallery \cite{kumarBricolageExamplebasedRetargeting2011}), or references from memory to earth animals or science fiction creatures when inventing new fictional alien creatures \cite{wardStructuredImaginationRole1994}). Importantly, examples} can substantially shape what ideas come to mind \cite{wardStructuredImaginationRole1994,smithConstrainingEffectsExamples1993}. \re{This \textbf{"structuring of imagination"} \cite{wardStructuredImaginationRole1994} is sometimes} helpful ``\textit{inspiration}" \re{that leads to more creative ideas}  \cite{eckertFortuneFavoursOnly1998,dahlInfluenceValueAnalogical2002,hargadonTechnologyBrokeringInnovation1997,siangliulue2015toward}. \re{But examples can also have harmful ``\textit{fixation}'' effects that constrain novelty and innovation} \cite{janssonDesignFixation1991,linseyStudyDesignFixation2010,wardStructuredImaginationRole1994,smithConstrainingEffectsExamples1993} (for recent reviews, see \cite{sioFixationInspirationMetaanalytic2015} and \cite{crillyWhereNextResearch2017}). \re{Importantly, examples can influence problem solving without conscious effort or recognition} \cite{marshInadvertentUsePrior1999,marshElicitingCryptomnesiaUnconscious1993,linseyStudyDesignFixation2010}, \re{and persist in spite of creators' explicit intentions not to be influenced by them} \cite{smithConstrainingEffectsExamples1993,vasconcelosEffectExplicitInstructions2018,janssonDesignFixation1991}. Perhaps in recognition of these facts, effective creators take an active role in finding, structuring and interacting with examples \cite{eckertFortuneFavoursOnly1998,herringGettingInspiredUnderstanding2009,sharminUnderstandingKnowledgeManagement2009b} using a variety of analog and digitally mediated systems and practices, such as search engines \cite{herringGettingInspiredUnderstanding2009,zhangInsituStudyInformation2020, palaniCoNotateSuggestingQueries2021, palaniActiveSearchHypothesis2021}, design workbooks and commonplace notebooks \cite{gaverMakingSpacesHow2011}, online whiteboards \cite{wallaceSketchyDrawingInspiration2020,macneilFreeformTemplatesCombining2023}, mood boards \cite{luceroFramingAligningParadoxing2012}, and wider interactions with their community of practice, such as trade publications and conventions \cite{eckertFortuneFavoursOnly1998,hargadonTechnologyBrokeringInnovation1997}.

% The core high level question we aim to answer in this work is: \textbf{what principled and systematic grounds should inform the design of interfaces for interacting with examples in the creative process?}
An important area of HCI research on creativity support tools \re{therefore} studies the design of interactive systems that can assist creators in discovering examples \cite{siangliulue2015toward,wangIdeaExpanderSupporting2010a,ivanovMoodCubesImmersiveSpaces2022,rhyscoxDirectedDiversityLeveraging2021a,wangInterpretableDirectedDiversity2022,kochSemanticCollageEnrichingDigital2020,kochImageSenseIntelligentCollaborative2020,brandtExamplecentricProgrammingIntegrating2010a, siangliulueProvidingTimelyExamples2015}, structuring, analyzing, and exploring collections of examples \cite{changRecipeScapeInteractiveTool2018,xuIdeateRelateExamplesGallery2021,siangliulueIdeaHoundImprovingLargescale2016,leeDesigningInteractiveExample2010,macneilProbMapAutomaticallyConstructing2021,changWebCrystalUnderstandingReusing2012}, %searching for  \re{\cite{kochSemanticCollageEnrichingDigital2020,kochImageSenseIntelligentCollaborative2020,brandtExamplecentricProgrammingIntegrating2010a, siangliulueProvidingTimelyExamples2015}}, analyzing examples \cite{changWebCrystalUnderstandingReusing2012}, 
and adapting and using examples \cite{ivanovMoodCubesImmersiveSpaces2022,kochSemanticCollageEnrichingDigital2020,kumarBricolageExamplebasedRetargeting2011}.
Designers of such systems need to grapple with an array of very practical interaction design decisions. For example, how should we support interaction with examples over different screen sizes? Should examples be delivered via recommendation (in small sets), a feed, or some other interaction paradigm? What information should be presented alongside an example? We would like to have a consistent theory to draw from to make these decisions. Beyond considerations of usability, we conjecture that such a theory would need to map design decisions (or classes thereof) to creativity-relevant behaviors and outcomes, ideally with a nuanced specification of the precise benefits and costs of each design decision for these behaviors and outcomes. A theory of \textbf{human-example interaction} like this could help us design better systems with sensible defaults, prioritize and negotiate design requirements, and guide evaluation.%what data we should collect to know if our system is helping or not. 

% Towards this end, our broader research goal is to construct a theory of \textbf{human-example interaction} that bridges rich theories of the psychological mechanisms of fixation and inspiration from examples, with the emerging body of ``strong concepts" \cite{hookStrongConceptsIntermediatelevel2012} and design patterns from HCI research on example-based creativity support systems. Our approach is to conceptualize and empirically validate mappings between design decisions --- represented at an appropriate level of description --- and psychological mechanisms of processing or using examples, which can then be mapped to creative outcomes in contextually appropriate ways.

As a step towards developing such a theory, we conducted an experiment with 182 participants solving a controlled analog to an exploratory creativity task \cite{bodenCreativeMindMyths2004}. %: we gave people the task of searching for ``rewards" in \re{an exploration space with a rugged and uncertain landscape (shown in previous work to benefit from more creative exploration vs. simple exploitation of local regions of the space \cite{masonCollaborativeLearningNetworks2012a})}. Participants were given a set of 10 examples (\re{10 locations in the exploration space, with} their corresponding values) as starting points\re{. We} 
We varied both the diversity of examples and the types of presentations: overlaid on the search environment (the ``\conIC" condition), presented in a list (the ``\conPar" condition), or in a dropdown selectable menu (the ``\conSeq" condition). The ``\conIC" design was inspired by an emerging pattern of \textit{contextualizing examples in the creator's workspace or problem} in HCI systems for example-based creativity \cite{sharminReflectionSpaceInteractiveVisualization2013,ivanovMoodCubesImmersiveSpaces2022,xuIdeateRelateExamplesGallery2021,webbFreeFormMediumCurating2016,kochImageSenseIntelligentCollaborative2020} on the one hand, and theoretical descriptions of the use of examples to (re)formulate problems \cite{helmsCompoundAnalogicalDesign2008,okadaAnalogicalModificationCreation2009,herringGettingInspiredUnderstanding2009,sharminUnderstandingKnowledgeManagement2009b,maierReasoningHumansII1931,kaplanSearchInsight1990}; the latter ``\conPar" and ``\conSeq" conditions were designed to be representative of common interfaces for interacting with examples (in search results lists and pages of recommendations).%We designed this task paradigm to give us a high degree of control over properties of the task (complexity, difficulty) as well as direct comparability between examples. 

Our primary results were threefold: 1) ``\conPar" presentation harmed solution quality compared with ``\conIC" or ``\conSeq" presentation; 2) each interface condition was associated with distinct self-reported example usage strategies (notably, more usage of examples to ``model" the problem space to guide exploration in the \conIC\space vs. \conPar\space or \conSeq\space conditions, and more usage of examples to ``stimulate" a specific direction of exploration in the \conPar\space condition); and 3) the \conPar\space condition's propensity for stimulation-based strategies was corroborated by an increased usage of ``hill-climbing" strategies early on, as evidenced by analyses of \conSeq\space distance between participants' moves. %These novel empirical results constitute the primary contribution of this paper. 

We discuss how these results, in conversation with the literature on example-based creativity support systems as well as psychological mechanisms of creativity with examples, could contribute to a theoretical framework for designing interactive systems for creative problem solving with examples. %Regarding the behaviors associated with each type of presentation, “\conIC” presentation of initial examples helped participants model the solution space, the “\conPar” presentation of initial examples stimulated participants by providing potential starting points; the “\conSeq” presentation will cause a barrier to the usage of the examples. 
% This initial framework constitutes the secondary contribution of this paper.

% These contributions advance our understanding of how to design creativity support tools. They may also advance fundamental understanding of how creativity works, since there are theoretical reasons to believe that situated theories of human cognition have better explanatory power than theories that attempt to abstract across interaction and situated contexts.

% To preview, our proposed theory articulates three main mechanisms by which diverse examples can support creative design: 1) encouraging \textbf{exploration} (and counteracting fixation), 2) enabling creative recombination and \textbf{integration} of disparate concepts, and 3) supporting more effective \textbf{modeling} of the design space. These mechanisms can be mapped to key interaction design decisions, such as \conSeq vs. \conPar presentation, and what metadata to include alongside examples.

% Design examples are an integral part of the creative design process \cite{buxton2010sketching, herring2009getting}. Exposure to examples might empower designers to be aware of potential options in the design space, and establishing connections between past solutions and the current context could inspire people creatively address new problems \cite{kim2005supporting, kolodner1993case}.

\section{Related Work}

% Our paper contributes new empirical insights into the effects of examples, new theoretical insights into human-example interaction, and new design implications for interactive systems for interacting with examples. To contextualize how our work builds on and contributes to these areas of research, we briefly review the literature on empirical variability in effects of examples, theoretical foundations of human-example interaction, and interactive systems for interacting with examples.

% \subsection{Examples as Double-edged Sword: Inspiration versus Fixation}
\subsection{Sources of empirical variability in effects of examples}
\label{sec:rw1}

Prior work has examined how the consequences of examples for creative problem solving outcomes are related to characteristics of the examples, such as their novelty \cite{agogueImpactTypeExamples2014,chanBenefitsPitfallsAnalogies2011,bergPrimalMarkHow2014,siangliulue2015toward} (generally positive effects), conceptual distance from the problem domain \cite{wardAnalogicalDistancePurpose1998,chanBestDesignIdeas2015,fuMeaningFarImpact2013,goncalvesInspirationPeakExploring2013,dahlInfluenceValueAnalogical2002} (mixed or curvilinear effects), and example diversity \cite{siangliulue2015toward,chanImportanceIterationCreative2015,zengFosteringCreativityProduct2011,howard-jonesSemanticDivergenceCreative2005,gielnikCreativityOpportunityIdentification2011,yuanExamplesCreativeExhaustion2022,doboliTwoExperimentalStudies2014} (generally positive or contingently positive effects). Our work contributes additional empirical results on the relative contributions of (and potential interactions, in the statistical sense, between) example characteristics and example \textit{interface} characteristics. In particular, we explore how the example characteristic of \textit{diversity} might interact with example interface characteristics, such as whether the examples are presented as a list vs. in context of a representation of the design space. To do this, we need to also consider the cognitive mechanisms of inspiration or fixation from examples (or varying characteristics), which might be more or less afforded by example interfaces. We discuss this body of literature in the next section.

\subsection{Theoretical insights into human-example interaction}
\label{sec:rw2}

% We also build on and contribute to examinations of how the effects of examples vary as a function of how creators \textit{process} them. For example, 
A number of detailed in-situ studies of creators have documented a range of strategies for working with examples, ranging from simpler, more source-driven strategies like direct source adaptation \cite{eckertAdaptationSourcesInspiration2003}, to more complex and reflective strategies associated with more radical transformation of \re{examples}, such as source analysis and schema-driven source selection \cite{eckertAdaptationSourcesInspiration2003}, analogical reasoning \cite{gentnerStructureMappingAnalogy1997,holyoakMentalLeapsAnalogy1996,gickSchemaInductionAnalogical1983,ballSpontaneousAnalogisingEngineering2004}, and generating novel emergent features that can connect disparate attributes across examples \cite{wilkenfeldSimilarityEmergenceConceptual2001}.
% in a think-aloud study of student and professional knitwear designers, Eckert and Stacey observed that the designers used a range of strategies for working with sources of inspiration, ranging from simpler, more source-driven strategies like direct source adaptation, to more complex and reflective strategies like source analysis and schema-driven source selection; complex strategies were more likely to be used by professionals and also more associated with radical transformation of sources \cite{eckertAdaptationSourcesInspiration2003}. Other areas of the literature have documented how using analogical reasoning \cite{gentnerStructureMappingAnalogy1997,holyoakMentalLeapsAnalogy1996} to abstract schemas from examples to guide their influence on ideation can be associated with more positive outcomes on creative outcomes \cite{gickSchemaInductionAnalogical1983,ballSpontaneousAnalogisingEngineering2004}, and how particular strategies of conceptual combination, such as generating novel emergent features that can connect disparate attributes of the examples, can facilitate more novel combinations of examples into new ideas \cite{wilkenfeldSimilarityEmergenceConceptual2001}. 
These ``processing strategies" can be described by a variety of theoretical frames from the psychological literature on creativity. We believe this theoretical level of description could facilitate our goal of synthesizing mappings between interface characteristics and effects of examples on creative problem solving outcomes. Some notable examples include basic memory mechanisms such as \textit{priming} \cite{mcnamaraPrimingConstraintsIt1992} and spreading activation \cite{collinsSpreadingactivationTheorySemantic1975,raaijmakersSearchAssociativeMemory1981}, and higher level cognitive processes such as conceptual \textit{abstraction} and \textit{analogical transfer} \cite{gickSchemaInductionAnalogical1983,dingFluidTransformersCreative2023}, \textit{conceptual combination} \cite{wilkenfeldSimilarityEmergenceConceptual2001}, and \textit{problem reformulation} based on examples \cite{helmsCompoundAnalogicalDesign2008,dorstCreativityDesignProcess2001,macneilProbMapAutomaticallyConstructing2021}. Of particular interest in our study is a contrast between priming and spreading activation mechanisms on the one hand, which is associated with lower-level conceptual influences, and problem (re)formulation, which is associated with  more complex, higher-order processing of examples. %Accordingly, we discuss these mechanisms in more detail here. 
In this study, we extend this literature by exploring how two specific mechanisms of processing examples (stimulation, and (re)formulation) might be helped or hindered by different example interaction interfaces. To set the context for our results, we briefly review the literature on each mechanism here.

\subsubsection{Using examples to stimulate ideation}
Spreading activation has been invoked to explain the impact of external stimuli on ideation. For instance, the ``search for ideas in associative memory" (SIAM) model  \cite{nijstadHowGroupAffects2006} %, which extends the classic ``search in associative memory" (SAM) model \cite{raaijmakersSearchAssociativeMemory1981}: in SIAM, 
proposes that when ideas come to mind, whether from memory, or through discussion with others or exposure to examples, they also raise the activation level of other associated concepts and features in memory, which can stimulate or inhibit ideation by making certain sets of ideas more or less likely to be generated based on the current network of associations in memory. For example, an example idea ``use as paperweight" (for a design prompt to generate alternative use for a brick) may activate related concepts such as ``office", or ``is heavy", along with their associated concepts; subsequent ideas such as ``construct a table", or ``prop up a bookshelf" may then be more likely to come to mind, compared to ideas like ``use as a weapon" or ``makeshift goalposts for soccer".
%proposes that when ideas come to mind, whether from memory, or through discussion with others or exposure to examples, they also raise the activation level of other associated concepts and features in memory. For example, the idea ``use as paperweight" (as an alternative use for a brick), may activate related concepts such as ``office", or ``is heavy", along with their associated concepts. Within this model, then, ideas from others or the world can stimulate or inhibit ideation by making certain sets of ideas more or less likely to be generated based on the current network of associations in memory. Returning to the brick as paperweight example idea, subsequent ideas such as ``construct a table", or ``prop up a bookshelf" will be more likely to come to mind, compared to ideas like ``use as a weapon" or ``makeshift goalposts for soccer". Thus, examples can be seen as influencing ideation by making certain subsequent ideas more or less likely, depending on the network of associations around that example. 
In this way, exposure to examples can shape the trajectory of ideation, and the corresponding floor or ceiling of creativity \cite{bergPrimalMarkHow2014,chanSemanticallyFarInspirations2017}. In this paper, we discuss this set of mechanisms under the label ``\textbf{stimulation}", to capture the intuition of examples stimulating ideation along a particular direction.%, though it should be noted that this set of mechanisms is frequently observed to have contingent/varying effects on creative outcomes, as exemplified by a number of ``fixation" and ``conformity" results discussed in the previous section. 

\subsubsection{Using examples to (re)formulate problems}
\label{sec:rw-modeling}
% A different area of the creativity literature also documents whether/how creators use examples to help them (re)formulate problems. 
Past research has documented how people can use examples to construct, refine, and even reformulate their understanding of the creative problem they are trying to solve \cite{helmsCompoundAnalogicalDesign2008,okadaAnalogicalModificationCreation2009,herringGettingInspiredUnderstanding2009,sharminUnderstandingKnowledgeManagement2009b,maierReasoningHumansII1931,kaplanSearchInsight1990}, through processes such as intentional free-form curation of examples \cite{lupferPatternsFreeformCuration2016} or on mood boards \cite{luceroFramingAligningParadoxing2012}. For instance, Okada et al. documented how two artists used individual artworks to shape not just their ideas, but used a process of ``analogical modification" to search for and modify higher-order concepts, and their creative vision over the course of years \cite{okadaAnalogicalModificationCreation2009}.
This process of example-influenced problem formulation is related to computational and neurobiological models of cognitive search \cite{hillsExplorationExploitationSpace2015} %, with broad applications across decision-making, reasoning, and problem-solving. The underlying question is 
which study \re{the range of strategies that intelligent agents can use to structure} their search processes: here, there is an important distinction between ``model-free" search, where external feedback from the world on the agents' actions guide search in a simpler, more local, stimulus-response manner, %(similar to the ``stimulation"-based mechanisms we briefly reviewed), 
and ``model-based" search, where the agent constructs a model or representation of the task and environment (in the case of creative problem-solving, this would be the problem and/or design space \cite{newellHumanProblemSolving1972,goelStructureDesignProblem1992,dorstCreativityDesignProcess2001}) \re{partly on the basis of reflection its own actions and possibly observation of others' actions}, and uses that model to decide where and when to explore in the task environment vs. continue to sample locally. Additionally, insight problem solving research has documented how people not only construct models, but also substantially revise them in radical ways, to solve difficult creative problems \cite{knoblichConstraintRelaxationChunk1999,kaplanSearchInsight1990}; this process is a key source of difficulty for creative problems, where one's initial problem formulation (e.g., key constraints or requirements), may be unhelpful \cite{knoblichConstraintRelaxationChunk1999}. 
In this paper, we discuss this set of mechanisms under the label ``\textbf{(re)formulation}", to capture the intuition of creators leveraging examples to (re)formulate their understanding of the design space. 

% \subsection{Interactions between Human and Examples}
\subsection{Interactive systems for interacting with examples}
\label{sec:rw3}

% As noted above, there is a large body of work in HCI exploring the design of interactive systems that assist creators in their efforts to benefit from examples. We are particularly 
In this study, we are interested in understanding how the impact of examples on problem solving varies as a function of interaction design decisions for how creators will interact with examples\re{.} 
% the connecting ``strong concepts" \cite{hookStrongConceptsIntermediatelevel2012} and design patterns \cite{linEmployingPatternsLayers2008} from HCI research on example-based creativity support systems to psychological mechanisms of fixation or inspiration. %Toward that end, we briefly review here examples of emerging design patterns in human-example interaction systems.
To set the context, we briefly review here some emerging interaction design patterns in HCI systems research into \textit{how} (vs. when, as in recommendation systems) participants interact with sets of examples (vs. understand or modify a single example).

One higher-level design pattern involves \textit{explorable overviews} of examples. For example, %Dream Lens is a visual analysis tool for modeling large-scale generative design datasets \cite{matejkaDreamLensExploration2018}; 
MetaMap \cite{kangMetaMapSupportingVisual2021} supports exploration of examples through keywords and colors and offering a playground to curate examples; RecipeScape \cite{changRecipeScapeInteractiveTool2018} uses a map UI to present recipe examples; Sifter \cite{pavelBrowsingAnalyzingCommandLevel} presents large collections of image manipulation tutorials in a faceted view based on their command-level structure; the Adaptive Ideas Web tool \cite{leeDesigningInteractiveExample2010} enables designers to explore and structure collections of web design examples; the Freed system \cite{mendelsFreedSystemCreating2011} empowers design students to spatially organize their digital collection of examples, define relations and reflect on their interrelationships; and Cabinet \cite{kellerCollectingCabinetHow2009} supports collecting and organizing of visual examples for inspiration and reference. 

Another emerging design pattern can be described as \textit{contextualizing examples in the creator's workspace or problem}, enabling designers to curate and reflect on the examples to build an understanding of their design space. %understand trends and design approaches to inform their design \cite{sharminReflectionSpaceInteractiveVisualization2013, sharminUnderstandingKnowledgeManagement2009b}. 
For example, ReflectionSpace \cite{sharminReflectionSpaceInteractiveVisualization2013} interactively contextualizes design artifacts in project timelines (and associated comments and reflections) to promote reflection and learning; MoodCubes \cite{ivanovMoodCubesImmersiveSpaces2022} enables designers to curate, compare, and explore suggested 3D design elements in the context of an overall 3D ``cube" room layout; %IdeaHound system uses implicit human actions with machine learning algorithm to generate a computational semantic model of a design space combining all solutions \cite{siangliulueIdeaHoundImprovingLargescale2016}. 
% \re{Side Views \cite{terrySideViewsPersistent} provides on-demand previews and transition examples of commands, overlaid on a text editor task context, to help users clarify, compare, and contrast commands.
IdeateRelate \cite{xuIdeateRelateExamplesGallery2021} locates design examples in coordinates of similarities corresponding to the users’ original ideas; the IdeaMache system provides an environment for free-form visual curation and sensemaking of creative materials in the context of a project canvas \cite{webbFreeFormMediumCurating2016}; and ImageSense embeds the process of searching for, exploring, and integrating examples into both individual and shared work spaces \cite{kochImageSenseIntelligentCollaborative2020}.

We build on this work by directly testing how the emerging pattern of contextualizing examples might impact the effects of examples on problem solving. To facilitate downstream theoretical development, we go beyond formulations of problem solving effects and outcomes that are task-specific --- such as writing code \cite{brandtExamplecentricProgrammingIntegrating2010a}, designing websites \cite{kumarBricolageExamplebasedRetargeting2011}, or designing room layouts \cite{ivanovMoodCubesImmersiveSpaces2022} --- and/or removed from creativity-specific mechanisms, such as browsing and searching and exploring, to more theoretically grounded descriptions of psychological mechanisms such as fixation and problem reformulation.

\section{Methods}
\label{sec:study1method}

% The analogy to creative tasks relies on a well-established underlying metaphor of searching in a space of solutions as a model for creative thinking \cite{newellHumanProblemSolving1972,hillsExplorationExploitationSpace2015}. The analogy does break down when considering, for example, 

% Here, we studied the effects of example characteristics and interface characteristics on this search process by givinzg all participants a set of 10 points in the grid as starting examples, varying both the diversity of examples (in terms of their spatial distribution over the 2D grid), and the interaction interface (how the examples were presented). %We were interested in exploring whether/how example interaction design choices might influence the degree to which people benefited from diverse examples. 
% We designed this task paradigm to give us a high degree of control over properties of the task (complexity, difficulty) as well as direct comparability between examples. 

% \subsection{Experimental Setup}

\subsection{The WildCat Wells Task as a Controlled Analog to Exploratory Creative Problem Solving}

We experimentally investigated our research questions using a controlled analog to \textbf{exploratory creativity}, a term introduced by Margaret Boden's influential model of creativity \cite{bodenCreativeMindMyths2004} to describe a subset of creative problem solving processes that involve \textit{exploration} within a conceptual space that is often large and complex. This conception of exploratory creative problem solving as search in a space has deep roots in research on search landscapes and innovation in organization and management science \cite{baumannEffectiveSearchRugged2019}, as well as psychological models of problem solving \cite{newellHumanProblemSolving1972} and creativity \cite{perkinsCreativityDarwinianParadigm1994} (as reviewed in our discussion of model-based mechanisms for using examples in \ref{sec:rw-modeling}). A key insight from this literature is that local search and hill-climbing are insufficient in more rugged and complex search landscapes, because they can trap searchers in local optima; to overcome this, searchers need to find ways to explore or ``jump" to new regions of the landscape \cite{baumannEffectiveSearchRugged2019}, such as guiding search through (re)modeling of the search space \cite{hillsExplorationExploitationSpace2015}. This contrast between local and distant exploration is often described in terms of the shift between exploitation and exploration \cite{baumannEffectiveSearchRugged2019}, where the latter search dynamics are more associated with innovation and creativity \cite{perkinsCreativityDarwinianParadigm1994}. Note that the notion of exploratory creativity is distinct from another important class of creative processes that involve what Boden \cite{bodenCreativeMindMyths2004} calls \textit{transformational} creativity: in this form of creativity, creators search for or construct alternative problem spaces (as discussed in the related work) \cite{dorstCreativityDesignProcess2001,kaplanSearchInsight1990}, rather than search within an existing problem space as given. 

Our controlled analog is the WildCat Wells task. The name of the task takes inspiration from the real-world task of wildcat drilling\footnote{https://en.wikipedia.org/wiki/Wildcatter}, a form of exploratory drilling for oil and gas in an unfamiliar environment where the distribution of resource-rich locations is unknown. \re{Accordingly, in this task, participants can ``drill'' for ``resources" in a 2D grid by clicking on locations in the grid; clicking on a grid location then uncovers a score amount, analogous to the amount of oil/gas uncovered at a drilling site.%, with approximate feedback on the score of ``objects" at each point. 
Like its real-world counterpart, the distribution of resources in this task is unknown; in our particular instantation, participants' goal is to uncover the most resource-rich drilling location (i.e., the grid location with the highest score).
Following our conceptualization of examples as descriptions/representations of possible solutions to the same/similar problem, in this task, we operationalized examples as possible grid locations and their associated scores.}

% Because the WildCat wells task is only analogous to exploratory creativity, our results here can only speak to effects of examples on exploratory, but not transformational, creative problem solving. 
\re{We chose this task for several reason. First, we had a high degree of parameter control over the properties of the task and examples, which allowed us to precisely control the ruggedness of the task structure and also employ a within-subjects design while mitigating learning effects by constructing and sampling from a set of Wildcat Wells' tasks with isomorphic ruggedness/complexity properties (see Section \ref{sec:task-search-env}). The task structure also gave us granular and precise measures of process and outcome dynamics. Finally, the simplicity of the task allowed us to minimize the impact of prior knowledge because the task does not require specialized domain expertise. While the specific task structure in terms of distribution of rewards over the search space is unknown to participants, the generic task structure of searching a space for rewards is probably not unfamiliar to most people. %, giving us comparability with previous results on problem solving. 
The Wildcat Wells task and its operationalization of examples is also conceptually similar to other instances of exploratory creativity that may draw on example solutions from a very similar problem: for instance, when searching for effective parameter settings for wing airfoil designs, other airfoil designs --- which, like our grid location, are also combinations of parameters --- may serve as relevant examples; when designing effective ads for a vaccine persuasion campaign, other vaccine persuasion ads --- which are also combinations of design features --- may serve as relevant examples; and when designing effective UI elements, other websites and their UI elements --- which are also compositions of UI features --- may serve as relevant examples. We also adapted the Wildcat Wells task specifically from a prior study \cite{masonCollaborativeLearningNetworks2012a} of the dynamics of exploration and exploitation in collaborative problem solving. However, b}ecause the WildCat wells task is only analogous to exploratory creativity, our results here can only speak to effects of examples on exploratory, but not transformational, creative problem solving.  %, and connects with the goal of larger research project where we aim to match human behavioral data with computational modeling. %Additionally, this was a relatively simple task (in terms of knowledge requirements), which we expected to reduce the influence of prior knowledge or verbal ability on performance. 

% Our Wildcat Wells experimental paradigm included two parts: a systematic means of 1) generating search environments with particular properties, and 2) selecting initial example sets of varying diversity. %Each combination of search environment and example sets constituted the materials for our experiment. 
% Our source code to generate the search environments and example sets is provided in the Supplementary Material I.

\subsubsection{Search Environments}
\label{sec:task-search-env}
\begin{figure*}
    \centering 
    \includegraphics[width=350pt]{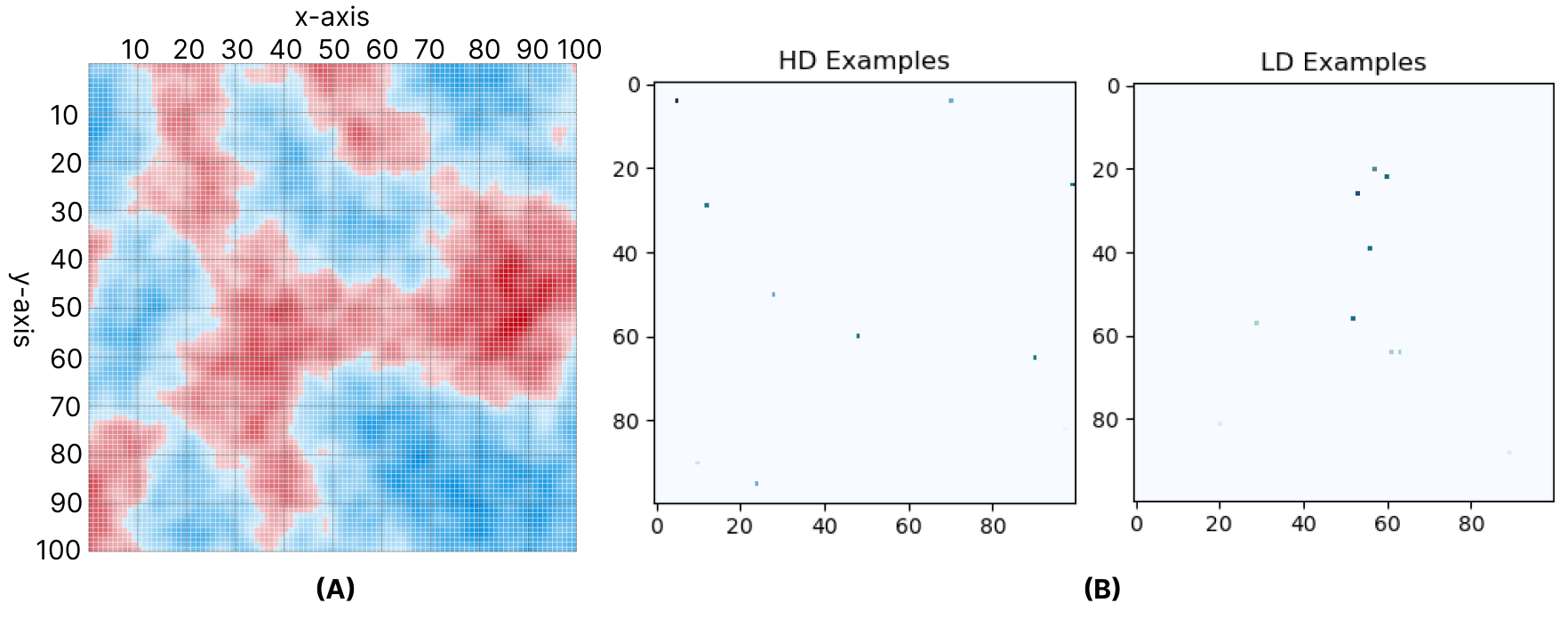}
    \caption{Example Wildcat Wells search environment with color coding of points to indicate their scores (0-50: dark blue to light blue, 50-100: light red to dark red) (A), and example sets of high and low diversity sets of points in this search environment, which are given as examples (B).}
    \label{fig:Land-HD-LD}
    \Description{TODO}
\end{figure*}

Our WildCat Wells search environments consisted of a 100x100 grid of points (with corresponding scores controlled by a synthetic objective function that determines the distribution of scores; see Algorithm 1 in Appendix \ref{sec:appendixA} and our source code for generating search environments). Figure \ref{fig:Land-HD-LD} (A) shows a representative search environment we used in our experiment.

% Following \cite{masonCollaborativeLearningNetworks2012a}, we constructed our synthetic objective functions to create a difficult version of this task, characterized by ruggedness and complexity. 
Our goal was to more closely match the difficult search spaces that the creativity theorist David Perkins calls ``Klondike spaces" \cite{perkinsCreativityDarwinianParadigm1994}, which are environments where simple ``hill-climbing" exploration strategies are insufficient, and likely outperformed by other creative exploration strategies such as a mix of exploration and exploitation \cite{perkinsCreativityDarwinianParadigm1994,baumannEffectiveSearchRugged2019}. We describe the specific parameter settings and algorithm we used to generate these task environments in Appendix \ref{sec:appendixA} (and share the code generating the environments in the Supplementary Material); here, we note that we set the parameters to yield a search environment that was fairly rugged (adding more false ``peaks" to incorrectly intuit as the location of the maximum \re{score}) and locally noisy (reducing the local correlation between scores in the grid, such that searchers would often be surprised by the score of nearby regions in the grid). The parameters to generate search environments were determined by a series of rubrics (e.g. more than one area with scores higher than 80), and pilots for a qualitative sense of difficulty based on the topology of the solution space. To reduce the likelihood that our results were tied to a specific formulation of the search environment, we generated ten search environments with the same synthetic objective function parameters but different random seeds. The resulting search environments were qualitatively similar to Mason \& Duncan’s work \cite{masonCollaborativeLearningNetworks2012a}. 
%smoothness level to 0.2 (fairly rugged), and the number of peaks to 1 with a maximum of 100; distance between peaks was not relevant because we only used a single peak. 

% Following \cite{masonCollaborativeLearningNetworks2012a}, we controlled the complexity and difficulty of the task by adding layers of Gaussian noise to add ruggedness to an otherwise smooth layer of normal distribution. 

\subsubsection{Examples}
\label{sec:task-examples}

To prepare sets of initial examples with \textbf{High Diversity} (HD) or \textbf{Low Diversity} (LD), we
% used the function below \todo{[function 1]} to 
randomly generated 10,000 sets of 10 examples each (recall that each example is \re{a "drilling location"} point in the 100x100 \re{Wildcat Wells} search environment, with a corresponding score) for each of our 10 search environments, and ranked the diversity of each example set with a close variant of the Determinantal Point Process (DPP) approach \cite{kulesza_determinantal_2012} described in Algorithm \ref{alg:dpp} in Appendix \ref{sec:appendixB} proposed by Eesh et al. \cite{kamrahHowDiverseInitial2023}; intuitively, this approach measured the ``hyper-volume" spanned by a selected set of points, such that larger volumes corresponded to higher levels of diversity, since these points spanned a larger set of the space of possible moves~\cite{kulesza_determinantal_2012}. We then randomly picked three HD examples sets that were greater than the 99th percentile of the distribution of diversity across the example sets, and three LD examples sets with diversity lower than 1st percentile of the diversity distribution. To ensure that examples would not directly reveal the location of the peak, or provide a high enough score that participants might simply stop after seeing the example instead of searching, we discarded example sets that had any point with a score over 80 (scores in the search environment ranged from 0 to 100), and resampled example sets as necessary --- subject to the same low/high diversity sampling criteria) to construct our final example sets for each search environment. Figure \ref{fig:Land-HD-LD} (B) shows an LD and HD example set used in our experiment.

% \todo{In Alg.~\ref{alg:dpp} $S^k$ is a combinatorial set defined on a finite set $X \in \mathbb{R}^2$, where each element $S^k_{Y_i} \in S^k$ is k elements long. Then, the cardinality of the set $|S^k|$ can be given by ${\prod |dim(X)}|\choose {k}$.}
%Lastly, the function $g$ mentioned in Alg.~\ref{alg:dpp} to calculate the DPP score has been adapted from \cite{faez ahmed and Mark}}.

%A list of N tests points was called a subset. A list of N subsets was called a set. Initially 10,000 of these subsets were generated randomly. Then we used a measure called Detrimental Point Processes (DPP) to give a diversity score to each subset. This measure calculated the negative log-determinant of the rbf kernel over the distance matrix generated for the test points in each subset. This gave a way to evaluate the volume spanned by these points. These subsets were then sorted relative to their scores. So, when this article refers to a subset with HD it means that the subset has been sampled from the 500 subsets with the highest diversity score in a set of these 10,000 subsets. Conversely, the subsets were chosen form bottom 500 subsets for LD.

%For the human experiment, we generated 10 100x100 spaces, and 3 HD and LD example sets for each space respectively. Each participant will be randomly assigned to two spaces, one space with one HD and one space with one LD example sets to mitigate the potential bias of one participant space and example set.

\subsection{Experiment Design}
\begin{figure*}
    \centering 
    \includegraphics[width=350pt]{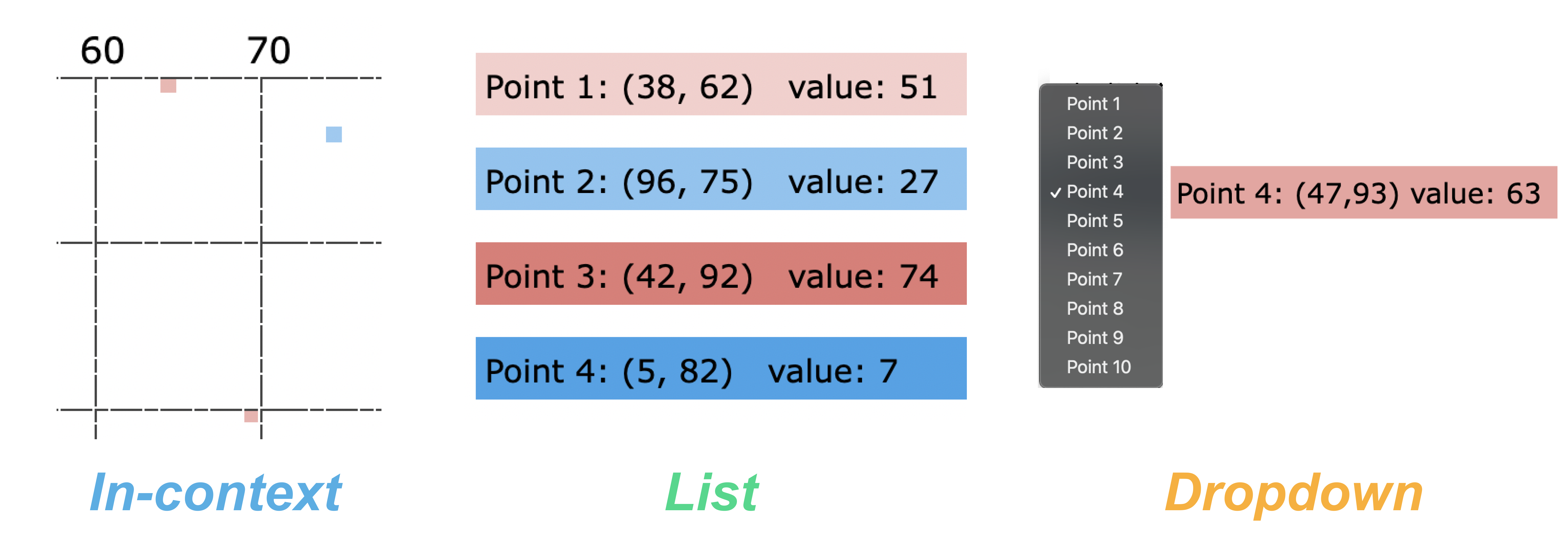}
    \caption{Three conditions of presenting examples: ``\conIC" (directly on the search environment grid), ``\conPar" (in a list) and ``\conSeq" (in a clickable dropdown selector).}
    \label{fig:3interfaces}
    \Description{TODO}
\end{figure*}

We conducted a mixed design experiment. \textbf{Example interface} was a \textit{between-subjects} factor, with three conditions: 1) ``parallel examples with context" interface: all 10 examples were shown in the 100x100 space with color coding to denote the score associated with each point, referred to as the ``\conIC" interface 2) ``parallel examples without context" interface (shown in Figure \ref{fig:Interface}): all 10 examples were shown in a list, also with color coding to indicate example score, referred to as the ``\conPar" interface 3) ``serial examples without context" interface: only one example was shown at a time and the participant needed to use a dropdown button to see other examples, referred to as the ``\conSeq" interface. Figure \ref{fig:Interface} shows the experimental interface of the \conPar\space condition as an example. The \conIC\space interface was inspired by design patterns of example interfaces that contextualized examples in the creator's workspace or problem (e.g., \cite{sharminReflectionSpaceInteractiveVisualization2013,ivanovMoodCubesImmersiveSpaces2022,webbFreeFormMediumCurating2016,kochImageSenseIntelligentCollaborative2020} %that integrated examples in a shared task or model environment that facilitated comparison and reasoning, such as \cite{changRecipeScapeInteractiveTool2018} and \cite{xuIdeateRelateExamplesGallery2021}. 
The \conPar\space interface was inspired by the familiar design pattern of a ``list" of examples, often in the context of a search interface (as search results), or list of recommendations in a recommender interface. The \conSeq\space interface was designed to approximate more constrained interfaces for interacting with examples, such as through chat-based or recommendation systems (e.g., popping up one or two examples at a time). The three example interfaces were shown in the context of the WildCat Wells task in Figure \ref{fig:3interfaces}. 
We conjectured that interfaces that allowed for comparison between examples (whether in the context of a task environment, as in the \conIC\space interface, or just with attributes shown for comparison, as in the \conPar\space interface) might facilitate more model-based usage of examples (what we called a ``(re)modeling" mechanism in \ref{sec:rw2}). Since we designed our Wildcat Wells task to be unsuitable for simpler hill-climbing %or ``gradient-descent" style strategies, where participants might pick a promising starting point from the list of examples and locally explore from there 
(e.g., ``stimulation-based" mechanism as described in \ref{sec:rw2}), we also expected that these interfaces might also lead to better performance on the task, through, for example, model-based exploration strategies. 

\textbf{Example diversity} was a \textit{within-subjects} factor: each participant attempted the WildCat Wells task twice, once with a set of HD examples, and once with LD examples. Recall that we generated 10 variant search environments, each with their own set of HD and LD example sets. To approximate counterbalancing of our within-subjects factor, we created 2 ``run" variants for each search environment, with each variant having an HD or LD example set as the first trial. Participants were randomly assigned first to an example interface condition, and then randomly assigned to one of the 20 potential ``runs" in each interface condition (but constraining assignment such that participants would not see the same search environment twice). %In this way, we approximated counterbalancing of example diversity order, as well as random assignment across search environments. 
Based on prior research on example diversity, we expected that participants would perform better when given high vs. low diversity example sets.

% \todo{The rationale for the conditions was that we expected the ``\conPar" and ``\conSeq" interfaces could stimulate the participants but not support participants' modeling behavior; only the ``\conIC" interface could support participants to ``model" the design space with examples.}

\subsection{Participants}
We recruited participants from the Amazon MTurk platform, limiting participants to U.S. residents with more than 500 HITs with at least 99\% approval rate. Each participant was paid US\$1.3 for their participation, which was an effective rate of \$10 per hour, given the average task completion time of 8 minutes. 

We aimed for a total sample size of 195 (65 per each of the three conditions), to achieve target statistical power of over 0.80 to detect medium-sized statistical effects in a mixed between-within design experiment analysis. %We rejected MTurkers if their responses in the survey were irrelevant, such as answering ``nice" to the survey question ``How did you use initial examples (the values of ten points given to you)?". 
After rejecting invalid work of 42 participants for irrelevant responses (e.g., ``nice") to the closing survey question about how they used examples), we obtained data from 182 participants (63 females, 118 males, 1 other; 65 in context, 56 \conPar, and 61 \conSeq) in total, yielding an effective statistical power of 0.86 for medium-sized effects. %There were 65 participants in the in context condition, 56 participants in the \conPar condition and 61 participants in the \conSeq condition. %From the post-study survey, all accepted participants had satisfactory task comprehension. A trial run was provided to make participants familiar and engaged with the task.

\subsection{Experimental Procedures}
\label{sec:procedure}
% should we mention the number of participants here? If so, mention numbers with or without rejection?
%\begin{quote}
%"We're interested in understanding how you usually solve problems creatively. In the next 2 minutes, please generate as many alternative uses as you can for the following common object: coffee cup. Please write each alternative use on a new line."
%\end{quote}

\begin{figure*}
    \centering 
    \includegraphics[width=300pt]{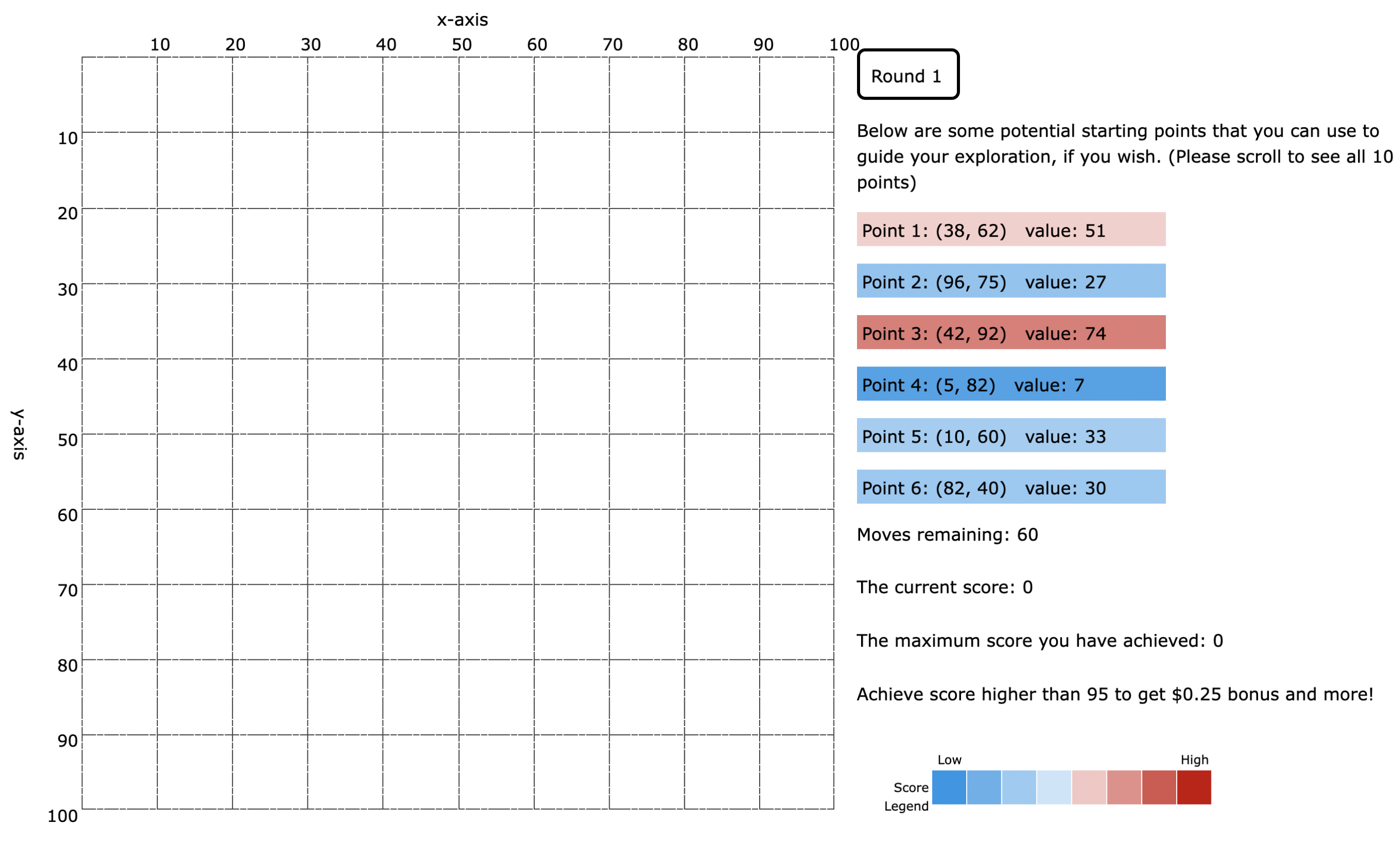}
    \caption{Screenshot of experimental interface, shown for the \conPar\space condition: the 100x100 grid, which constituted the search environment for the task, was shown on the left panel: participants explored the space by clicking anywhere on the 100x100 grid. The 10 initial examples, moves remaining, the score of current move, the current max score and score legend were shown on the right panel. In the \conSeq\space condition, the dropdown menu as seen in Figure \ref{fig:3interfaces} was shown in the same position as the list of examples in the \conPar\space condition. In the \conIC\space condition, examples were instead overlaid as points, with corresponding values, on the search grid, as shown in Figure \ref{fig:3interfaces}.}
    \label{fig:Interface}
    \Description{TODO}
\end{figure*}

Participants experienced the WildCat Wells task as a 100x100 space (see Figure \ref{fig:Interface}). Their task was to find the square with the highest score. Participants explored squares by clicking on them to reveal their underlying score, shown in color coding, similar to the examples. %The overall distribution of scores was otherwise hidden from the participants. 
To simulate the constrained nature of real creative tasks (which often have some time/budget pressure) and reduce the likelihood of ceiling effects, participants had a total budget of 60 moves for exploring squares. This budget was estimated from our pilot studies, where on average, most participants found the highest scoring square within 50 moves. We also provided incentives to encourage participants' exploration: there was a \$0.25 bonus for achieving a highest score greater than 95, and a \$0.50 bonus for achieving the maximum score of 100. The information panel on the right side of the experimental interface (see Figure \ref{fig:Interface} showed moves remaining, the score of the current exploration and the maximum score the participant had achieved in the current round.

Since the WildCat Wells task does not interact strongly with prior knowledge in any particular domain, %we reasoned that background information such as background professions might not be informative as control variables. Instead, 
we addressed potential pre-existing differences in ability by measuring participants' baseline divergent thinking ability, a correlate of creative ability \cite{runcoDivergentThinkingCreativity2010}. %Divergent thinking --- the capacity to fluently and flexibly generate many different alternave solutions to a problem --- is a robustly studied psychological correlate of creative ability; though it is not synonymous with creativity, the psychological literature has consistently found moderate (coefficients in the range of .3 to .5) correlations between divergent thinking ability and creative achievement \cite{runcoDivergentThinkingCreativity2010}. Before the study, 
Before the study, we asked participants to generate as many alternative uses of coffee cup as they could in 2 minutes (an instance of the commonly used Alternative Uses task \cite{guilfordNatureHumanIntelligence1967} for measuring divergent thinking \cite{runcoDivergentThinkingCreativity2010}). Participants were then given one trial round through the WildCat Wells task (without examples) to familiarize them with the interface and task. After that, participants completed two formal rounds of the WildCat Wells task, which constituted the main experimental trials in our study. %The trial round was designed to measure a baseline of participants' exploration ability without initial examples and also serve as a practice task to familiarize the participant with the study interface. In those formal rounds, participants would be given some initial information about squares on the space to guide their exploration. Those two formal rounds had an example set with LD and an example set with HD respectively. The order of low and high diversities was balanced. Each round had 60 moves for participants to explore.
Finally, participants completed a post-study questionnaire, with three free-response questions: 1) What strategy did you use for hunting? 2) How did you use initial examples (the values of ten points given to you)? 3) What differences did you notice between initial examples given in those two rounds? Which did you find helpful?

We obtained institutional IRB approval for the whole project prior to the study.

% \subsection{Measures}

% The primary measures we used for our analysis were 1) performance and sequence data such as the locations and values of participants' moves, and clicks of the dropdown menu in the \conSeq condition (as a rough indicator of interactions with examples); and 2) participants' response to the post-study survey question \textit{"How did you use initial examples (the values of ten points given to you)?"}. We also used the number of responses on the pre-study alternative uses task to estimate whether random assignment across between-subjects factors succeeded with respect to divergent thinking ability. 

\section{Results: Planned Analyses}

\subsection{No significant differences in baseline divergent thinking ability across interface conditions}

% Before we get into the main results from our planned analyses, 
\re{We first report the results of our check for random assignment with respect to divergent thinking ability and baseline performance on our task.}
We observed no statistically significant difference in the number of generated alternative uses across three conditions ("\conIC" participants: $M=6.52, SD=3.02$; ``\conPar" participants: $M=5.75, SD=3.58$; ``\conSeq" participants: $M=6.46, SD=3.87$, Kruskal-Wallis $H=2.40, p=0.30$). 
\re{Similarly, we observed no statistically significant difference in participants' best score on the trial run of the Wildcat Wells task across the conditions ("\conIC" participants: $M=90.97, SD=7.52$; ``\conPar" participants: $M=90.91, SD=7.13$; ``\conSeq" participants: $M=91.18, SD=7.37$, Kruskal-Wallis $H=0.19, p=0.91$). This suggests that participants across the interface conditions were comparable in terms of baseline divergent thinking ability as well as baseline task performance.}
% Figure \ref{fig:explore_example} shows the illustration of one round in this study.

% \begin{figure}
%     \centering 
%     \includegraphics[width=350pt]{figures/Explore-example.jpg}
%     \caption{An illustration of the study. Left: the  landscape of the score distribution (the darkness implies the value); Middle: the initial HD examples presented to the participant; Right: the locations of the participant's 60 moves.}
%     \label{fig:explore_example}
%     \Description{TODO}
% \end{figure}

% \subsection{\conPar Presentation of Examples Was Associated with Worse Design Performance}

\begin{figure*}
    \centering 
    \includegraphics[width=0.9\linewidth]{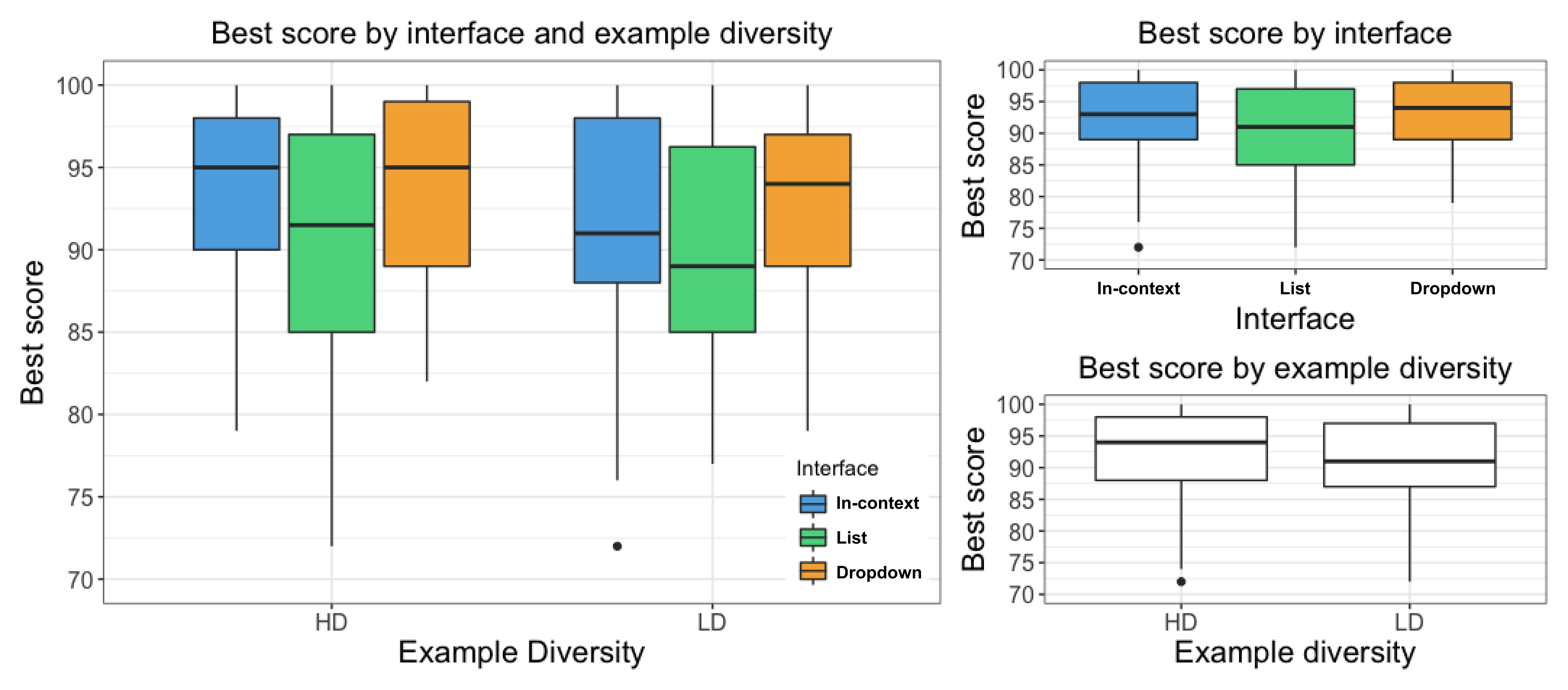}
    \caption{Distribution of best scores by interface and example diversity conditions. Participants in the \conPar\space interface condition had lower best scores than participants in the other interface conditions regardless of example diversity (top right). Best scores were also lower when participants were given low vs. high diversity examples (bottom right).} %The error bars are standard error of proportion.}
    \label{fig:max_distribution_boxplot}
    \Description{Figure 4 presents the distribution of best scores by interface and example diversity conditions with three boxplots. Participants in the \conPar\space interface condition had lower best scores than participants in the other interface conditions regardless of example diversity (top right). Best scores were also lower when participants were given low vs. high diversity examples (bottom right).}
\end{figure*}

\subsection{\conPar\space presentation of examples and low diversity example sets associated with lower best scores}
\label{sec:res1}

% \subsection{\conIC Interface Associated with Better Design Performance, regardless of Example Diversity}

% \begin{table}[b]
% \begin{tabular}{|l|l|l|l|l|l|}
% \hline
% \textbf{Con} & \textbf{Total} & \textbf{HD = 100} & \textbf{HD >= 95} & \textbf{LD = 100} & \textbf{LD >= 95} \\ \hline
% In context & 65 & 13 (20.0\%) & 34 (52.3\%) & 11 (16.9\%) & 26 (40.0\%) \\ \hline
% \conPar & 56 & 6 (10.7\%) & 19 (33.9\%) & 8 (14.3\%) & 17 (30.4\%) \\ \hline
% \conSeq & 61 & 15 (24.6\%) & 31 (50.8\%) & 7 (11.5\%) & 26 (42.6\%) \\ \hline
% \end{tabular}
% \caption{Counts of participants achieving scores equal to 100 and larger or equal than 95 with the \conIC and \conPar interfaces.}
% \label{tab:max_count}
% \end{table}
% Next, we report our main set of planned analysis: overall associations between interface condition and example diversity and task performance. 
% Figure \ref{fig:max_distribution_boxplot} plots raw ``best scores" (the highest score achieved by participants in each trial) across the interface and example diversity conditions. 
% We observe that most participants performed quite well (overall median between 90 and 95), though there was substantial variability across participants (see Fig. \ref{fig:max_distribution_boxplot}). 
The \conPar\space condition had slightly lower scores on average compared to the other conditions (regardless of example diversity; Fig. \ref{fig:max_distribution_boxplot}, top right). There was also an overall slight advantage of HD examples over LD examples (Fig. \ref{fig:max_distribution_boxplot}, bottom right). %To statistically examine these associations between our interface and example diversity manipulations and task performance, we first estimated 
A linear mixed effects model with best score as the dependent variable,  interface condition and example diversity as factors, and random intercepts for participants (estimated in the `lme4` package in `R`), showed a significant main effect of interface condition, F(2, 179) = 5.92, $p < .01$, $eta^2$ = 0.06. Pairwise post-hoc comparisons with Bonferroni corrections showed that participants in the \conPar\space interface condition had significantly lower best scores (est. marginal mean = 90.4, SE = 0.65) compared to both the \conIC\space (est. marginal mean = 92.7, SE = 0.61, contrast t ratio = 2.54, $p < .05$) and  \conSeq\space interface conditions (est. marginal mean = 93.4, SE = 0.63, contrast t ratio = 3.31, $p < .01$). There was also a significant main effect of example diversity, F(1, 181) = 4.59, $p < .05$, $eta^2$ = 0.02; post-hoc comparisons with Bonferroni corrections showed that participants had higher best scores when they received HD (est. marginal mean = 92.7, SE = 0.46) vs. LD examples (est. marginal mean = 91.5, SE = 0.46, contrast t ratio = 2.14, $p < .05$). 
%examined how these factors were related to the highest score that participants achieved by the end of each trial; two thresholds were relevant, given our task design: how many participants achieved scores equal to or larger than 95 (which was tied to a \$0.25 bonus) and equal to the global maximum of 100 (tied to a \$0.50 bonus). Then, we examined how these factors were related to the average score achieved across each condition. \todo{Figure \ref{fig:max_distribution} plots these metrics across the conditions.}

%The proportion above each score threshold across the conditions is shown in Table \ref{tab:max_count} and Figure \ref{fig:max_distribution}. Since we had three interface conditions, we applied Kruskal-Wallis H-test, a rank-based non-parametric test for more than two groups on the max scores of those three conditions with high and low diverse examples respectively. There was a significant difference among those three conditions with high diverse examples ($statistic=9.57, p-value=0.008$). While there was no significant difference among those three conditions with low diverse examples ($statistic=4.02, p-value=0.13$).

\section{Results: Exploratory Analyses}
\label{sec:res_explore}

We conducted a set of exploratory analyses to better understand the results of our main planned analyses, focusing on understanding process effects of interface conditions that might plausibly explain performance differences. %1) a deeper dive into performance differences over time as a function of interface condition, 2) associations between interface condition and self-reported example strategy use, and 3) associations between interface condition and early exploration behavior.

\subsection{\conIC\space presentation of examples associated with early performance advantages, and \conPar\space presentation of examples with early and persistent performance disadvantages}
\label{sec:res_explore1}

\begin{figure*}
    \centering
    \includegraphics[width=0.8\textwidth]{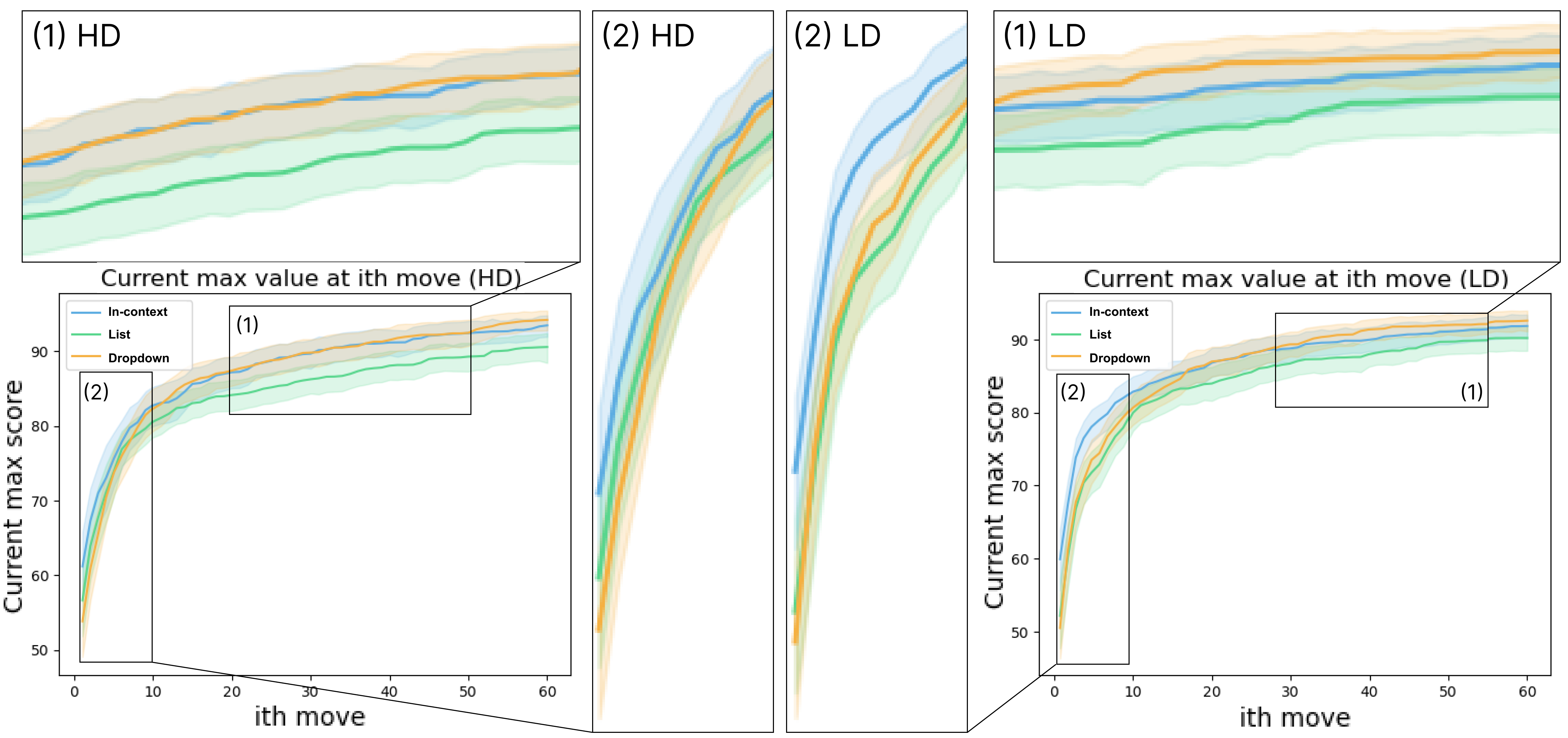}
    \caption{Maximum score at n-th move for each participant: left) HD; right) LD. We observe (1) cumulative disadvantages for the \conPar\space condition, as well as (2) early advantages for the \conIC\space interface, especially with LD examples.}
    \label{fig:max_step}
\end{figure*}

First, for a more granular view of performance, we examined how the participants' best score changed as a function of their move sequence. This analysis confirmed a cumulative disadvantage for participants in the \conPar\space condition, but also showed an early advantage for the \conIC\space interface, particularly with LD examples (see Figure \ref{fig:max_step}). Using a Kruskal-Wallis H-test on the current max score from the 1st to the 30th move, we observed statistically significant differences from the 1st move to the 6th move and the 8th move except the 7th move (see Table \ref{tab:move_diff}).

% \begin{figure}
%      \centering
%      \begin{subfigure}[b]{0.49\textwidth}
%          \centering
%          \includegraphics[height=0.6\textwidth]{figures/HD_progress.png}
%          \caption{}
%     \label{fig:hc_all}
%      \end{subfigure}
%      \hspace{-4em}%
%      \begin{subfigure}[b]{0.5\textwidth}
%          \centering
%       \includegraphics[height=0.6\textwidth]{figures/LD_progress.png}
%          \caption{}
%     \label{fig:hc_split}
%      \end{subfigure}
%         \caption{Maximum score at n-th move for each participant: a) HD; b) LD. We observe cumulative disadvantages for the \conPar condition, as well as early advantages for the \conIC interface, especially with LD examples.}
%       \label{fig:max_step}
% \end{figure}

\begin{table*}%[b]
    \begin{tabular}{|l|l|l|l|l|l|}
        \hline \textbf{\textit{n}-th move} & \textbf{\conIC\space ($M$)} & \textbf{\conPar\space ($M$)} & \textbf{\conSeq\space ($M$)} & \textbf{Statistic} & \textbf{p-value} \\ \hline
        1 & 60.06 & 52.30 & 50.64 & 9.13 & 0.01 \\ \hline
        2 & 67.65 & 60.45 & 61.39 & 6.93 & 0.03 \\ \hline
        3 & 74.00 & 66.95 & 67.89 & 9.58 & 0.008 \\ \hline
        4 & 76.65 & 70.54 & 70.84 & 9.06 & 0.01 \\ \hline
        5 & 78.20 & 71.93 & 73.66 & 10.13 & 0.006 \\ \hline
        6 & 79.15 & 73.07 & 74.59 & 9.59 & 0.008 \\ \hline
        7 & 79.98 & 75.07 & 76.87 & 5.93 & 0.06 \\ \hline
        8 & 81.46 & 76.86 & 78.21 & 6.91 & 0.03 \\ \hline
    \end{tabular}
    \caption{Means ($M$) and results of Kruskal-Wallis H-test on the current best score for three interface conditions with LD examples. There were statistically significant differences between the conditions from the 1st to 8th moves (p<0.05 except the 7th move).}
    \label{tab:move_diff}
\end{table*}

%This advantage of the \conIC interface was especially prominent early on with LD examples as shown in Figure \ref{fig:max_step}.  \todo{To investigate the difference between HD and LD examples within each interface condition, we applied Mann–Whitney U test, a rank-based non-parametric test on the max scores with HD and LD examples for each interface condition respectively, and there was no significant difference: \conIC ($statistic=1841.0, p-value=0.10$), \conPar ($statistic=1503.0, p-value=0.35$), \conSeq ($statistic=1548.5, p-value=0.05$).}

% \begin{figure}
%     \centering 
%     \includegraphics[width=180pt]{figures/100-95.png}
%     \caption{Less ``\conPar" participants' max scores achieved the 100 and 95 thresholds with examples in general. The error bars are standard error of proportion.}
%     \label{fig:max_distribution}
%     \Description{TODO}
% \end{figure}

% \begin{figure}
%     \centering 
%     \includegraphics[width=180pt]{figures/performance_split_3.png}
%     \caption{"\conPar" participants had lower best scores than participants in the other interface conditions regardless of example diversity. Best scores were also lower when participants were given low vs. high diversity examples. The error bars are standard error of proportion.}
%     \label{fig:max_distribution}
%     \Description{TODO}
% \end{figure}

\subsection{Variations in example presentation interfaces associated with different self-reported example usage strategies}

% \subsection{Variations in Presentation Interfaces Were Associated with Different Self-Reported Example Usage Strategies}

% \subsubsection{"\conIC" participants self-reported more \model\space usage of examples; ``\conPar" participants self-reported more \stimulate\space usage of examples; over half of ``\conSeq" participants did not use examples.}

Next, to understand how participants used the initial examples in their exploration, two researchers coded participants' responses to the question ``How did you use initial examples (the values of ten points given to you)?" with three codes: \textbf{not using}, \textbf{\stimulate} and \textbf{\model}. This classification was guided and refined by our initial theoretical interest in the contrast between stimulation-based and (re)modeling-based use of examples, as discussed in \ref{sec:rw2}. Examples of responses coded as ``not using" include \textit{"I did not give much thought to it"}, and \textit{"Not much to be honest"}; examples of ``stimulation-based" responses included \textit{"Start at the reddest one and explore its surroundings"}, and \textit{"I looked around the higher values for boxes that were darker"}; examples of ``model-based" responses include \textit{"To get an overview on which squares would be best"}, and \textit{"They gave a vague idea of whether or not there might be ``hot" or ``cool" zones around those points"}. When we could not infer how the participants used the initial examples, the answers were coded as ``unclear". The researchers were blinded to condition during coding. Inter-rater reliability was substantial, at Cohen's $\kappa$ = 0.725  \cite{landisMeasurementObserverAgreement1977a}; all disagreements were resolved by discussion. %Examples of each code are provided in Table \ref{tab:code_example}. 

\begin{figure}
    \centering 
    \includegraphics[width=220pt]{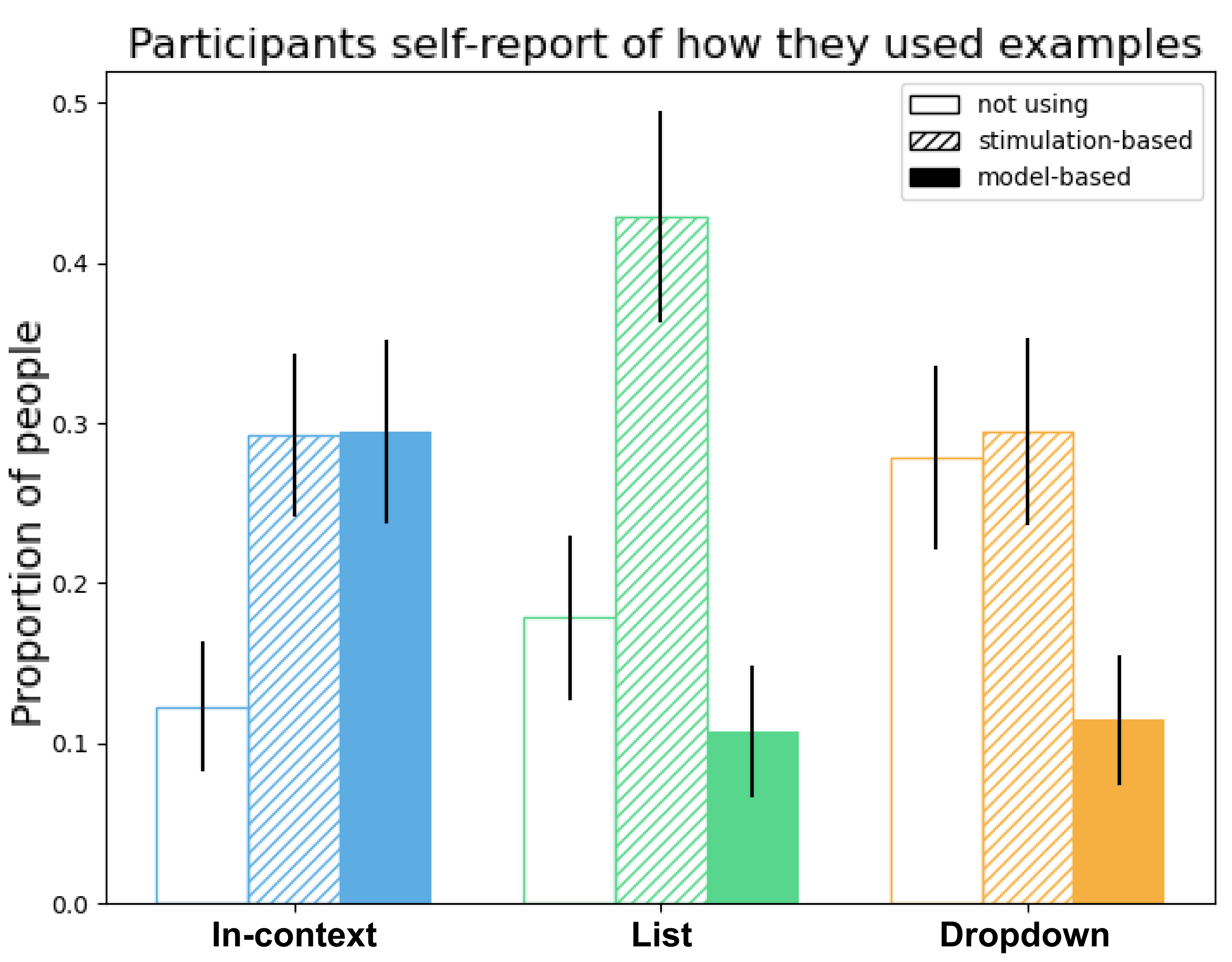}
    \caption{Raw proportion of participants expressed ``not using (examples)", ``\stimulate" or ``\model" in their answer to ``How did you use initial examples (the values of ten points given to you)?". Error bars are standard error of proportion. More participants self-reported using a model-based strategy in the \conIC\space condition compared to other conditions.}
    \label{fig:example-usage}
    \Description{TODO}
\end{figure}

% Figure \ref{fig:example-usage} shows the raw proportions (with standard errors of proportion) of coded survey responses for ``not using", ``stimulate", and ``model", across the conditions. 
The \conIC\space condition had the largest portion (30.8\%) of participants who self-reported using the initial examples to \textit{model} the space, compared to the \conPar\space condition (10.7\%) and the \conSeq\space condition (11.5\%) (see Figure \ref{fig:example-usage}). In contrast, for self-reported use of initial examples to \textit{stimulate} their exploration, the \conPar\space condition had the highest percentage (42.9\%) followed by the \conSeq\space condition (29.5\%) and the in context condition (29.2\%). Finally, 17/61 (27.9\%) \conSeq\space participants self-reported that they were not using examples, which was higher than participants in the other two conditions. Our log data were consistent with this observation: %in the \conSeq\space condition, the participants could click the dropdown button to see initial examples provided; however, 
37/61 (60.7\%) participants in the \conSeq\space condition never clicked the dropdown button to see other examples in both HD and LD conditions. Of the remaining participants who did click on the dropdown button, we infer --- assuming that each subsequent click corresponds to an example view --- that the mean number of examples viewed was M = 7.37 (SD = 3.40) for HD, and M = 7.05 (SD = 3.61) for LD. %Table \ref{tab:sequential_click} presents the statistics of number of clicks of dropdown button made by and number of examples seen by participants in the \conSeq condition.

% \begin{table}%[b]
%     \begin{tabular}{|l|l|l|l|l|l|}
%         \hline  & \textbf{Clicks made (HD)} & \textbf{Clicks made (LD)} & \textbf{Examples seen (HD)} & \textbf{Examples seen (LD)} \\ \hline
%         \textit{mean} & 2.721 & 2.918 & 2.984 & 3.082  \\ \hline
%         \textit{std} & 5.067 & 5.398 & 3.509 & 3.570  \\ \hline
%         \textit{min} & 0 & 0 & 1 & 1  \\ \hline
%         \textit{median} & 0 & 0 & 1 & 1  \\ \hline
%         \textit{max} & 17 & 20 & 10 & 10 \\ \hline
%     \end{tabular}
%     \caption{Number of clicks of dropdown button made by and number of examples viewed by participants in the \conSeq condition. Half of the participants never clicked the dropdown button and only saw the initial example.}
%     \label{tab:sequential_click}
% \end{table}}

\subsubsection{\conIC\space participants more likely than other interface conditions to self-report model-based example usage, and \conSeq\space participants more likely than other conditions to self-report not-using examples.}
We first statistically tested these patterns with a series of logistic regressions, one for each example strategy (not-using, \stimulate, and \model) (see Table \ref{tab:survey_code}). We ran separate logistic regressions rather than a single multinomial regression given our interest at this step in the relative likelihood across interface conditions of self-reporting a particular example strategy, rather than relative differences across strategies within each condition (best answered by a multinomial logistic regression). We first observe that participants in the \conPar\space and \conSeq\space conditions were less likely to self-report using a model-based strategy compared to the \conIC\space condition ($B$ = -1.31, 95\% CI = [-2.31, -0.31], z = -2.57, $p$ < .05 for \conPar\space vs. \conIC, and $B$ = -1.23, 95\% CI = [-2.18, -0.29], z = -2.55, $p$ < .01 for \conSeq\space vs. \conIC). In more intuitive terms, 
% \re{C}onverting these log odds coefficients to more intuitive terms to odds ratios, these estimates can be interpreted as 
\conIC\space participants were approximately 3x more likely to self-report using a model-based example usage strategy compared to \conPar\space or \conSeq\space participants (Odds Ratio = 3.7 and 3.4 for \conIC\space vs. \conPar, and \conIC\space vs. \conSeq). The overall model fit was better than a null model ($LL_{model}$ = -80.93 vs. $LL_{null}$ = -86.16), Likelihood Ratio $\chi^2$(2) = 5.23, $p < .01$. Next, we observe that participants in the \conIC\space condition were less likely to self-report not using examples compared to the \conSeq\space condition ($B$ = -1.01, 95\% CI = [-1.94, -0.09], z = 2.14, $p$ < .05); in more intuitive odds ratio terms, %the odds ratio for \conSeq vs. \conIC participants' odds for this contrast was approximately 2.75, which means that 
\conSeq\space participants were ~2.7x more likely than \conIC\space participants to self-report a ``not using" strategy (Odds Ratio = 2.75). The overall model fit, though better than a null model ($LL_{model}$ = -86.62 vs. $LL_{null}$ = -89.10), was marginally significant, Likelihood Ratio $\chi^2$(2) = 5.14, $p = .08$. Finally, we observe that there were no significant differences across conditions in the likelihood of self-reporting a stimulation-based strategy. Indeed, the overall model fit, though nominally better than the null model ($LL_{model} = -114.52$ vs. $LL_{null} = -116.08$ was not statistically significant, Likelihood Ratio $\chi^2$(2) = 1.56, $p = .21$.

\subsubsection{\conPar\space participants more likely to self-report a stimulation-based example usage strategy compared to not-using or model-based example usage.}
%\todo{Add in multinomial logistic regression to test predominance of stimulation-based strategy for \conPar}
Next, we focus on statistically evaluating the apparent predominance of a stimulation-based self-reported example usage strategy for \conPar\space participants. We fitted a multinomial logistic regression, with model-based usage as the reference outcome class. Participants in the \conPar\space condition were significantly more likely to self-report using a stimulation-based strategy compared to a model-based one, $B = -1.44$, 95\% CI = [-0.34, -2.53], z = -2.58, $p < .01$. In odds ratio terms, %This coefficient can be translated to an Odds Ratio of 4.21 for this contrast, which means that 
participants in the \conPar\space interface condition were ~4x more likely to self-report a stimulation-based vs. model-based strategy (Odds Ratio = 4.21). The overall model fit was statistically significantly better than a null model, Likelihood Ratio $\chi^2$(4) = 14.39, $p < .01$.

\begin{table*}%[b]
    \begin{tabular}{llll}
    \toprule
    &\textbf{Not using} & \textbf{Stimulation-based} & \textbf{Model-based} \\ \bottomrule
    Intercept & -0.951\: [-1.511, -0.391] & -0.288 [-0.817, 0.242] & \:-0.811 [-1.338, -0.284] \\ %\hline
    \conIC & \textbf{-1.013} [-1.940, -0.085]* & -0.597 [-1.349, 0.156] &  \\ %\hline
    \conPar & -0.575\: [-1.459, \:0.309] &  & \textbf{-1.309} [-2.307, -0.312]** \\ %\hline
    \conSeq &  & -0.583 [-1.347, 0.180] &  \textbf{-1.232} [-2.179, -0.285]* \\ \bottomrule
    \end{tabular}
    \caption{Coefficient estimates from multinomial logistic regressions of probability of self-reported example usage strategy (1 each for not-using, \stimulate, \model) on interface condition. Statistics reported as \textit{"coefficient, 95\% CI ([lower, upper])}". When the cell for a given interface condition is blank, that condition was used as the reference class in the regression. *p < .05; **p < .01; ***p < .001.}
    \label{tab:survey_code}
\end{table*}

% Due to that the participants in the \conSeq condition frequently ignored the examples, we dropped the \conSeq condition in the main analysis to and included the \conSeq condition in the appendix to guarantee the transparency of our study result.

\subsection{\conPar\space presentation of examples associated with more local initial exploration of the solution space}
% \subsection{\conPar Presentation of Examples Was Associated with More Local Initial Exploration of the Design Space}
% \subsubsection{More “\conPar” participants did hillclimbing for the first 30 moves with LD examples than other combinations of the presentation and example sets.}

Finally, we explored how log data might be consistent (or not) with participants' self-reported example usage strategies. We wanted to study how initial examples would affect participants' exploration behaviors, especially at the beginning of exploration when the examples provided were a major source of information. To explore this, we first constructed an exploration graph for the first 30 moves of each participant trial by computing Euclidean distances between each successive move; the intuition was that long sequences of low distances between moves would suggest ``hill-climbing", and large distances would suggest ``jumps". We conjectured that a ``hill-climbing"-like exploration graph would be consistent with a stimulation-based strategy, rather than a model-based strategy. 

\begin{figure*}
    \centering 
    \includegraphics[width=370pt]{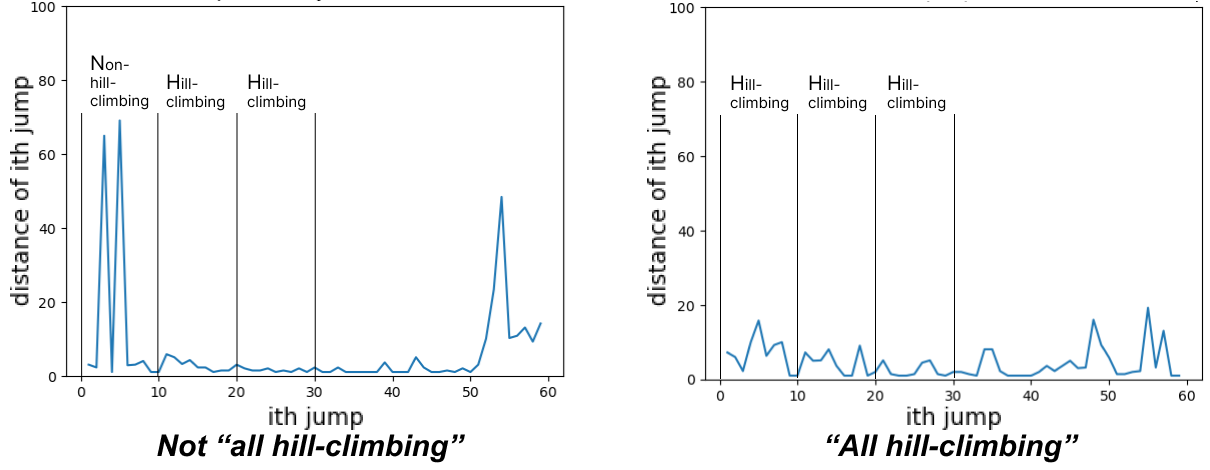}
    \caption{Example exploration graphs used for coding hill-climbing strategies: not ``all hill-climbing" (left) and ``all hill-climbing" (right), where each transition between participant moves is plotted on the x-axis, and the Euclidean distance between each move and its immediately preceding move is plotted on the y-axis. In the not ``all hill-climbing" example, the first 10 moves (the 0th-10th moves), where there are substantial variations in \conSeq\space move distances across the sequences, would be coded as non-hillclimbing (N), while the 10th-20th moves and the 20th-30th moves, where \conSeq\space move distances are consistently low, would be coded as hillclimbing (H). In the ``all hill-climbing" example, all first 30 moves would be coded as hillclimbing (H).}
    \label{fig:coding}
    \Description{TODO}
\end{figure*}

Two coders independently coded all 364 exploration graphs (each of the 182 participants had an HD plot and a LD plot), coding whether the exploration behaviors were hillclimbing (h) or not (n) for sequences of 10 moves (the 0th-10th, the 10th-20th, the 20-30th). Two examples of coding are shown in Figure \ref{fig:coding}. We coded 1092 10-move instances (3 10-move instances per round x 2 rounds per participant x 182 participants) with substantial inter-rater reliability, Cohen's $\kappa$ = 0.78.

\begin{figure}
     \centering
     \begin{subfigure}[b]{0.45\textwidth}
         \centering
         \includegraphics[height=0.6\textwidth]{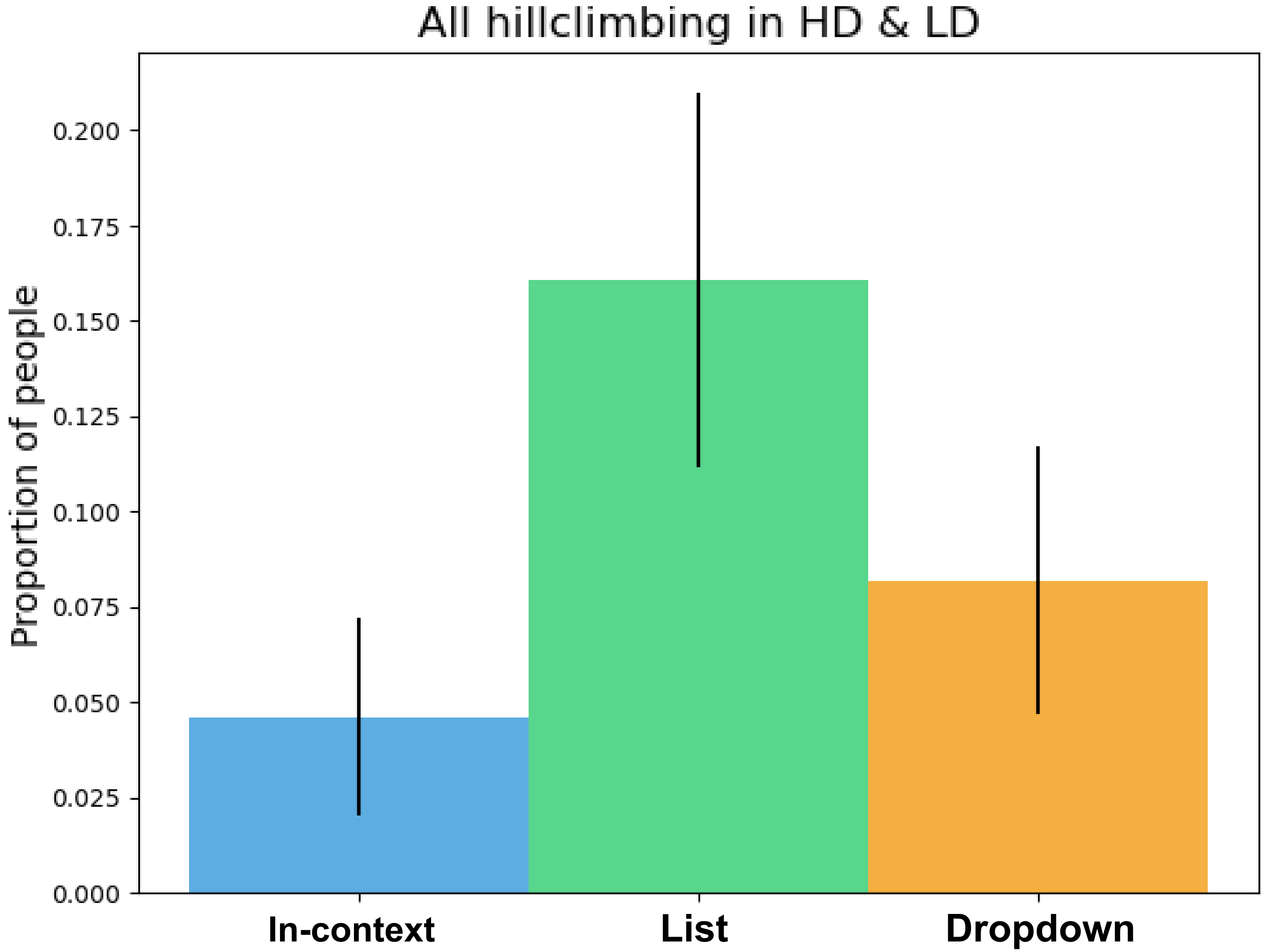}
         \caption{}
    \label{fig:hc_all}
     \end{subfigure}
     \begin{subfigure}[b]{0.45\textwidth}
         \centering
      \includegraphics[height=0.6\textwidth]{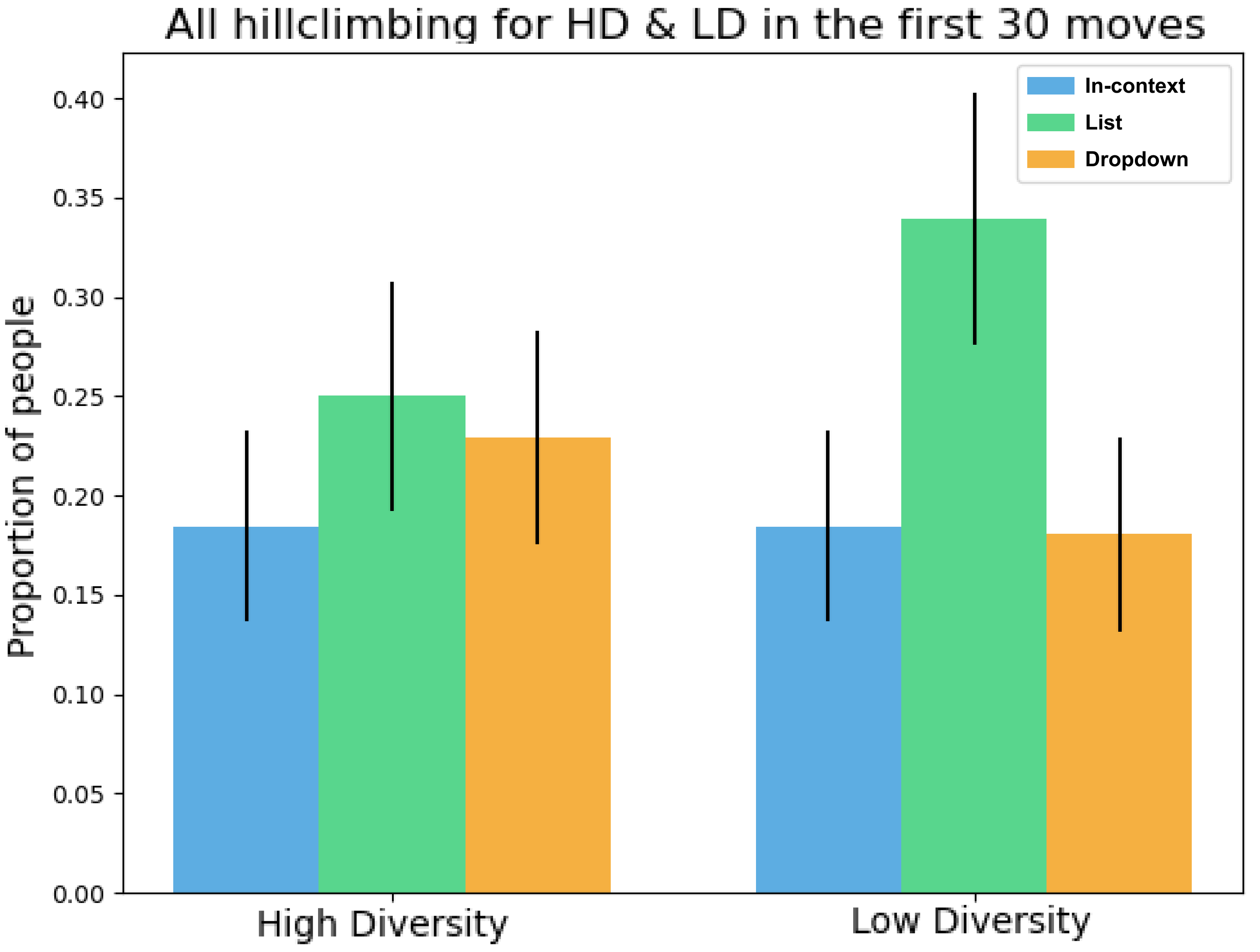}
         \caption{}
    \label{fig:hc_split}
     \end{subfigure}
       \caption{Raw proportion of participants with a predominantly hillclimbing strategy in the first 30 moves, across interface conditions (a) with both HD and LD examples (b). More ``\conPar" participants did hillclimbing in the first 30 moves with both high and low diverse example sets than ``\conIC" participants. More “\conPar” participants did hillclimbing for the first 30 moves with LD examples than other combinations of the presentation and example sets.}
      \label{fig:hill_climbing}
\end{figure}

% We were most interested in whether/how interface conditions would be associated with a predominantly hill-climbing strategy. %(in particular, whether the \conPar interface would be associated with this strategy, consistent with the survey results). Figure \ref{fig:hill_climbing} shows the raw proportions of participants in each condition who were coded as displaying a hill-climbing strategy in their first 30 moves, for HD and LD trials separately (B) and aggregated (A). 
For all trials, a higher proportion of participants in the \conPar\space interface condition used an exclusively hill-climbing strategy for the first 30 moves (proportion=0.161, lower bound=0.112, upper bound=0.210) compared to the \conIC\space (proportion=0.046, lower bound=0.020, upper bound=0.072) and \conSeq\space condition (proportion=0.082, lower bound=0.047, upper bound=0.117; see Figure \ref{fig:hc_all}a). Similarly, for LD trials, the proportion of participants in using an exclusively hill-climbing strategy for the first 30 moves was higher for the \conPar\space condition (34\%) compared to the \conIC\space (18\%) and \conSeq\space condition (18\%) (Fig. \ref{fig:hc_all}b, left). In contrast, for HD trials, the proportion of participants using an exclusively hill-climbing strategy for the first 30 moves was similar for the \conIC\space (18\%), \conPar\space (25\%), and \conSeq\space conditions (23\%) (Fig. \ref{fig:hc_all}b, left).%We observe that more participants (34\%) did hillclimbing for the first 30 moves in the \conPar interface with LD examples than other combinations of the interface and example sets (18-25\%).

To statistically test these observations, we fitted a series of logistic regressions, estimated with maximum likelihood, predicting $p(all\_hill)$, the probability of being all hill-climbing in the first 30 moves as a function of $interface$. Prior work suggests that choice of exploration vs. exploitation %(which often manifests as hill-climbing behavior) 
is influenced by the ``goodness" of the current region of the search space (better scores makes hill-climbing more likely) \cite{hillsExplorationExploitationSpace2015,baumannEffectiveSearchRugged2019}. Our data confirmed this pattern: the average score of the first move in each of the first three 10-move blocks was positively correlated with the likelihood of being all hill-climbing in the first 30 moves, Kendall's $\tau$ = .17, $p < .01$. Thus, we conditioned our logistic regression models on the average score at the beginning of each 10-move block.

We first analyzed $p(all\_hill)$ aggregated across both HD and LD trials (value would be 1 if both LD and HD trials were 1), and then HD and LD trials separately.
% all hillclimbing in both HD and LD
Table \ref{tab:hill_climb_models} shows the coefficient estimates for each of these models. For \textbf{all trials}, the coefficient for the contrast between the \conPar\space and \conIC\space conditions was $B$ = -1.79, 95\% CI=[-3.40, -0.45], z = -2.45, $p < .05$; in odds ratio terms, %This coefficient translates to an odds ratio in favor of the \conPar condition, $1 \div \exp^{-1.79} = 5.99$, indicating that 
participants in the \conPar\space condition were ~6x more likely to use an exclusively hill-climbing strategy for the first 30 moves, compared to participants in the \conIC\space condition (Odds Ratio = 5.99). Note that this effect was independent of the significant positive coefficient for the average first score in the block. The overall model fit was statistically significantly better than a null model ($LL_{model}$ = -48.42 vs. $LL_{null}$ = -56.48), Likelihood Ratio test  $\chi^2$(3) = 16.12, $p < .01$.
% hillclimbing in either HD or LD
Similarly, for \textbf{LD trials}, 
there was a statistically significant coefficient in the logistic regression model for the contrast between the \conIC\space and \conPar\space conditions, $B$ = -0.95, 95\% CI=[-1.83, -0.10], z = -2.16, $p < .05$. In odds ratio terms, %This coefficient translates to an odds ratio in favor of the \conPar condition, $1 \div \exp^{-0.95} = 2.58$, indicating that 
participants in the \conPar\space condition were ~2.5x more likely to use an exclusively hill-climbing strategy for the first 30 moves, compared to participants in the \conIC\space condition (Odds Ratio = 2.58). There was a similar contrast between the \conSeq\space and \conPar\space conditions, $B$ = -0.95, 95\% CI=[-1.85, -0.10], z = -2.12, $p < .05$. As with the all trials model, this effect was independent of the significant positive coefficient for the average first score in the block. The overall model fit was statistically significantly better than a null model ($LL_{model} = 92.98$ vs. $LL_{null} = -98.32$), Likelihood Ratio test $\chi^2$(3) = 10.67, $p < .05$. In contrast, there were no statistically significant contrasts between conditions in the \textbf{HD trials}, though the numerical pattern of results were similar to the other models (generally negative coefficients for \conIC\space and \conSeq\space vs. \conPar\space conditions). The overall model fit, though nominally better than a null model ($LL_{model} = -95.14$ vs. $LL_{null} = -95.85$, was not statistically significant, Likelihood Ratio $\chi^2$(3) = 1.42, $p = .70$.

\begin{table*}%[b]
    \begin{tabular}{llll}
    \toprule
    &\textbf{All trials} & \textbf{HD trials} & \textbf{LD trials} \\ \bottomrule
    Intercept & \:-7.322 [-11.73, -3.739] & -1.734 [-3.522, -0.052] & -2.551 \:[-4.409, -0.870] \\ %\hline
    \conIC\space vs. \conPar & \textbf{-1.789} [-3.400, -0.449]* & -0.431 [-1.325, \:\:0.448] & \textbf{-0.950} [-1.833, -0.103]* \\ %\hline
    \conSeq\space vs. \conPar & \:-0.926 [-2.220, \:0.260] & -0.113 [-1.325, \:\:0.448] & \textbf{-0.946} [-1.850, -0.0858]** \\ %\hline
    Avg. first score in block & \:\:\textbf{0.089} [\:0.035, \:0.152]** & \:\:0.010 [-0.016, \:\:0.037] &  \:\:\textbf{0.031} [\:0.005, \:0.059]* \\ \bottomrule
    \end{tabular}
    \caption{Coefficient estimates from logistic regressions of probability of predominantly hill-climbing strategy in the first 30 moves on interface condition and average first score in block, across all, HD, and LD trials. Statistics reported as \textit{"coefficient, 95\% CI ([lower, upper])}". *p < .05; **p < .01; ***p < .001.}
    \label{tab:hill_climb_models}
\end{table*}

Because we were concerned this pattern of differences might be driven by pre-existing individual differences in propensity to hill-climbing, rather than a shift due to the interface condition, we repeated our coding procedure for exploration graphs generated from the initial trial round of exploration, which did not include examples (described in \ref{sec:procedure}. The proportion of participants who displayed a predominant hill-climbing strategy, as described above, was distributed across conditions as follows: \conIC = 0.05 (SE = .03), \conPar = .07 (SE = .03), and \conSeq = 0.15 (SE = .05). A logistic regression predicting the probability of a predominant hill-climbing strategy (yes or no) as a function of interface did not improve fit over a null model with no predictors, $\chi^2$ (2) = 4.17, $p=.12$; note, however that the overall frequency of hill-climbing strategies was lower than in the main trials, and the \conPar\space condition was not the condition with highest frequency (in contrast to the main trials). There was also no significant correlation between the likelihood of hill-climbing predominance and either the number of alternative uses task responses, r = .04, $p = .61$, or the likelihood of hill-climbing predominance in the main trials, r = .03, $p = .65$. Altogether, these results are inconsistent with the alternative explanation that \conPar\space participants were simply more likely (due to individual differences) to choose a predominantly hill-climbing strategy overall; instead, taken together with the survey results, we believe this set of results suggest a shift in strategy towards hill-climbing (or, as we describe in this paper, a stimulation-based strategy for using examples).

\section{Discussion and Conclusion}
\subsection{Summary and Interpretation of Results}
In this paper, we aimed to contribute to a theory of human-example interaction to guide the design of example-based creativity support tools. %set out to explore what an interaction-oriented theory of creative inspiration from examples might look like. 
Towards this goal, we conducted an experiment with a controlled analog to an exploratory creativity \cite{bodenCreativeMindMyths2004} task to investigate how example presentation interface variations influence whether/how people benefit from examples. %We now summarize and interpret the results of our planned and exploratory analyses with regard to these questions. 

% First, 
We found evidence that \textbf{\conPar\space presentation of examples might harm the quality of final solutions}. For example we found variations across interface conditions in mean best score obtained at the end of trials across example interface and diversity conditions (\conPar\space participants had worse best scores compared to \conIC\space and \conSeq\space participants; Section \ref{sec:res1}) and cumulative performance differences (with an early and persistent disadvantage of the \conPar\space condition compared to the other conditions, with an especially pronounced early disadvantage relative to the \conIC\space condition; Section \ref{sec:res_explore1}). This result is conceptually significant because the “\conPar” participants received more information (seeing all 10 examples at the same time) than the “\conSeq” participants (only seeing 1 example at a time. and over 60\% of them never checked other examples), and approximately equivalent information but different presentation compared with the “\conIC” condition. This suggests that seemingly unimportant, low-level interaction design decisions with respect to presentation of examples can have measurable consequences for creative problem solving performance. Separately, we also observed \textbf{beneficial effects of diversity for final solution quality}, in line with some previous work \cite{siangliulue2015toward,baruahCategoryAssignmentRelatedness2011,zengFosteringCreativityProduct2011,howard-jonesSemanticDivergenceCreative2005,gielnikCreativityOpportunityIdentification2011,althuizenSupportingCreativeProblem2014}; importantly, this effect was similar in magnitude to the example presentation effects, suggesting that example presentation considerations may be just as important to consider as example characteristics when designing example-based creativity support systems.

Second, our exploratory analyses suggest that \textbf{"\conIC" and ``\conPar" presentation of examples may lead to distinct patterns of example usage}. “\conIC” presentation of initial examples was associated with a greater likelihood of a ``model-based" strategy for using examples, where participants self-reported using the examples to gain an overall understanding of the distribution of score in the search environment to guide their exploration, compared to \conPar\space or \conSeq\space presentations. Conversely, the “\conPar” presentation of initial examples seemed to encourage a predominant ``stimulation-based" example usage strategy, where participants selected promising examples as starting points for their exploration. Importantly this self-report data was consistent with patterns in our log data: we observed that \conPar\space participants were more likely to use a predominantly ``hill-climbing" strategy (with low \conSeq\space distance between their moves) early in their exploration, relative to the \conIC\space and \conSeq\space participants; this association was independent of the relationship between hill-climbing behavior and the ``goodness" of initial moves (hill-climbing in a given block of moves was more likely when the initial move was higher-scoring, consistent with prior empirical work on exploration/exploitation decisions \cite{hillsExplorationExploitationSpace2015,baumannEffectiveSearchRugged2019}). %We noticed that the goodness of the score of the first move in a block was positively related to the likelihood of hill-climbing. However, this would not change the pattern of results regarding the difference between \conIC and \conPar participants, because \conIC participants actually had higher initial scores compared to the \conPar participants, as seen in Figure \ref{fig:max_step}; they would therefore be more likely to hill-climb than the \conPar participants, on the basis of the goodness of their scores. So it strengthened the contrast between \conIC and \conPar conditions on hill-climbing intent.} 
Considering these results alongside the performance results suggests that \conPar\space participants were being \textit{fixated} \cite{janssonDesignFixation1991} by the examples. 

\re{A fruitful direction for further research would be to investigate the mechanisms that drive fixation in the \conPar\space condition.} One reason might be the upper limit in scores (no more than 80/100) on the examples presented to the participants; if taken as starting points to begin hill-climbing, those relatively low quality examples could be misleading, and block access to high-quality solutions. Another reason might be the increased effort needed to connect examples of the \conPar\space condition to the search space, which would be consistent with past research on %well-known mechanisms in the cognitive science of diagrammatic representations, including 
the cognitive load benefits of integrating diagrams and text (similar to integrating examples and the search space) in instructional design \cite{chandlerSplitAttentionEffectFactor1992}. \re{We view the difficulty of transferring from the text modality of lists of examples to the visuospatial modality of the \conIC\space solution space as a potential \textit{mechanism} by which example presentation variations might shape their impact on ideation: future work could investigate in more detail how different example presentation designs might shift the \textit{cost structure} of different processing strategies, in a similar way that variations in environment or interface structure have been shown to shape sensemaking by changing the cost structure of various crucial actions, such as skimming/previewing, moving documents, applying schemas to documents, or adjusting schemas \cite{russellBeingLiterateLarge2006,pirolliInformationForaging1999,russellCostStructureSensemaking1993}.}

Separately, we observed that the “\conSeq” presentation was associated with limited usage of the examples: many \conSeq\space participants self-reported not using examples (moreso than \conIC\space participants, for example), and this was also corroborated in their log data (via a lack of interaction with the example interface). \re{We do not think that this lack of example usage is indicative of a lack of engagement: recall, for instance, that performance in this condition was on par with the \conIC\space condition (i.e., higher than in the \conPar\space condition). Post-survey comments indicating enjoyment and engagement (e.g., "Fun game. Thank you!") were also seen across conditions at similar rates, and there were no statistically significant mean differences across the conditions in the trial run of the task. For these reasons, we believe that --- possibly due to the interaction affordances ---} the \conSeq\space condition appeared to act similarly to a "no-examples" control condition, where participants used a wider mix of strategies vs. a particular set of example-based strategies tied to an experimental intervention. In light of this, the overall strong performance of the \conSeq\space condition is akin to past observations of strong performance by control ``no-intervention" conditions in ideation experiments (see, e.g., \cite{chanBenefitsPitfallsAnalogies2011, siangliulue2015toward}; we thus add to a growing body of evidence that it may be easier to harm rather than help creative ideation by intervening (as in the \conPar\space condition). 

Overall, our results suggest that interaction design considerations for human-example interaction go beyond usability: there is indeed a space of mappings to explore between design affordances and fundamental psychological mechanisms of creative inspiration from examples.
%The behaviors of using examples are as expected since for most participants, modeling the 100x100 space with initial examples required a clear presentation of the scores and locations of initial examples, which a list of the XY axis presentation failed to satisfy. The advantage of model-based behavior over stimulate-based behavior is that modeling behavior could provide hints on which part of the solution space is more promising, meanwhile mitigating the fixation effect caused by the initial examples.
From a practical standpoint, our empirical results suggest the limitations of only showing examples without the problem space as context, especially if the problem space is large (a common feature of real world problems) and there exist some potential solutions far away from the initial examples. This implication is significant since the “\conPar” view of examples - examples presented in a list - is commonly used in current creativity support tools, such as search engines and recommendation systems, yet was associated with substantial negative effects on the usage of examples and task performance relative to \conIC\space presentation of examples. %Finally, showing examples sequentially might be problematic if the goal is to maximize the impact of (well-chosen) examples. 

\subsection{Limitations}
% We next discuss some limitations and scope considerations of our study before concluding with implications for our larger goal of building a theoretical bridge between interaction design considerations and fundamental psychological mechanisms of human-example interaction.

The WildCat Wells task we used in our experiment is simpler than most real-world exploratory creativity tasks\re{ --- such as airfoil design, ad design, or UI design ---} of which it is an analog. \re{For instance, the task did not require any specialized domain knowledge, and the generic task structure of searching a space for rewards is probably familiar to most people: indeed, one participant in our study noted that in the post-survey that the task was "very fun and somewhat similar to minesweeper." Additionally, }although we carefully constructed our Wildcat Wells task surfaces to be rugged, with multiple peaks of good solutions, our task technically has a single best solution; in contrast, many real-world creative problems \re{--- such as policy design ---} %are ``wicked problems" \cite{rittelDilemmasGeneralTheory1973}, 
lack a single best solution, \re{due to task factors such as intrinsic tradeoffs between different problem requirements; in these cases, creators often} search for and construct ``good enough" solutions under high uncertainty \re{(though this might sometimes be a function of feasibility constraints rather than intrinsic properties of the task). %This simplicity allowed us to minimize the impact of prior knowledge, and feasibly employ a within-subjects design with granular process and outcome measures, 
% \re{but warrants caution when generalizing to other more complex instances of exploratory creativity.
It is unsurprising, then, that} 
% One clear difference is that participants in this study could make 60 moves in a few minutes, in contrast to the longer timescales for iterating on problem solving or design moves in real-world settings. Additionally, real-world creative problem solving tasks often require more domain expertise than this exploratory problem solving task. Hence, it is unsurprising that 
participants performed relatively well as a whole, %: the maximum scores of 16.5\% (60/364) rounds could reach the highest score of 10000 points within 60 moves, and the maximum scores of 42.0\% (153/364) rounds could get equal or larger than 95, the top 1\%-7\% scores of 10000 points within 60 moves. 
and “\conSeq” participants also had competitive performance even though they did not interact or use the examples. \re{We note, however, that performance was not quite at ceiling: only 16\% of participants reached the global max in either trial (and 42\% reached the threshold score of 95 for the first bonus. Still, caution is warranted when generalizing to other more complex instances of exploratory creativity; for instance, i}t may be that the effects of examples, and the corresponding effects of variations in their presentation interactions, will become more pronounced in more sophisticated tasks. 

% For instance, participants were able to complete the tasks quite quickly. 
% Additionally, 
 
\re{Relatedly, the Wildcat Wells task} captures aspects of search dynamics (exploration and exploitation) in exploratory creative problem solving quite well, but does not enable observation of more sophisticated psychological mechanisms \re{for working with examples. For instance}, it is unclear what it might mean to ``combine" different problem solving moves for this task. Additionally, while participants engaged in modeling of the problem space, they were not able to make larger changes to the problem space, such as questioning assumptions or relaxing constraints \cite{knoblichConstraintRelaxationChunk1999}, or even changing the goal/problem altogether \cite{kaplanSearchInsight1990}, mechanisms that are common in real-world creative problem solving tasks, such as design \cite{dorstCreativityDesignProcess2001}. Thus, we reiterate that our results cannot speak to how example presentation design decisions might influence example usage for transformational creativity \cite{bodenCreativeMindMyths2004} tasks. 

Thus, more work is needed to extend our exploration of patterns in example interaction design choices to more complex settings: for example, what might it mean to design a ``contextualized" presentation of examples for \re{UI elements, more complex airfoil designs, ad persuasion campaigns,} research papers, or policy ideas? We are keen to build on existing design patterns similar to this in previous systems such as ReflectionSpace \cite{sharminReflectionSpaceInteractiveVisualization2013}, MoodCubes \cite{ivanovMoodCubesImmersiveSpaces2022}, and ImageSense \cite{kochImageSenseIntelligentCollaborative2020}, as discussed in Section \ref{sec:rw3}. Our implementation of the \conSeq\space condition may also be quite different from other \conSeq\space presentations, such as forward/backward interfaces (e.g., image suggestions \cite{kochMayAIDesign2019}). Future studies can explore the consequences of these differences. For now, we note that our main results on the contrast between \conIC\space and \conPar\space conditions are independent of this limitation, and recommend caution in generalizing the results from the \conSeq\space condition around non-use of examples.

Finally, %although we provided monetary incentives to crowdworkers for achieving higher scores, added a trial round to increase their task comprehension and engage them in the task, and rejected participants without thoughtful survey response, it is possible that some participants may have minimized effort/attention in the task in order to get the base payment quickly. Separately, 
% we were able to measure and control for possible baseline differences in divergent thinking ability \cite{guilfordNatureHumanIntelligence1967} as a plausible control variable for the relatively low knowledge requirements WildCat Wells task. However, 
we did not measure demographic information that may have been correlated with task performance or example usage and/or exploration patterns -- for example, personality traits such as disagreeableness or extraversion may be correlated with real-world creative achievement \cite{zabelinaCreativeAchievementIndividual2021}; and gender might interact with potential differences in visuospatial reasoning demands between example interfaces, given some existing research on %small but consistent 
gender differences in spatial ability \cite{lauerDevelopmentGenderDifferences2019}.%, such as mental rotation 

\subsection{Towards an interaction-oriented theory of creative inspiration from examples} 
Returning to our higher-level goal of constructing an interaction-oriented theory of human-example interaction, we now reflect on how the empirical results from our study, in conversation with theoretical mechanisms and design patterns from prior work, could contribute to an overall theory that bridges design patterns to psychological mechanisms.% in a way that enables reasoning about connections to creative processes and outcomes.

We conjecture that a useful theory of human-example interaction could be conceptualized as paths through multiple coordinated spaces of \textbf{example interaction patterns}, \textbf{example-ideation psychological mechanisms}, \textbf{ideation characteristics} and \textbf{creative outcomes}. Paths through this overall set of coordinated spaces could then represent a set of principled design hypotheses about how to best support creative work with examples. 
%Importantly, there may be multiple such paths for different task environments. Figure \ref{fig:theorysketch} depicts one such possible path, 
For instance, bringing our empirical results in conversation with the literature we reviewed in sections \ref{sec:rw2} and \ref{sec:rw3}, we could hypothesize that, given a particular exploratory creativity task environment like our instantiation of the WildCat wells task, where the key \textit{creative outcome} of \textbf{solution quality} is determined at least in part by the \textit{ideation characteristic} of \textbf{diversity of search}, which is in turn positively influenced by the \textit{psychological mechanism} of \textbf{(re)modeling}, and negatively influenced by the mechanism of \textbf{stimulation}, it may be advantageous to choose \textit{example interaction patterns} like \textbf{contextualizing examples in the problem space} (which is positively mapped to (re)modeling mechanisms), over patterns like \textbf{\conPar\space viewing of examples} (which is positively mapped to stimulation mechanisms). Multiple other hypothesized paths could be generated and refined to map other example interaction patterns from prior work, such as faceted search systems, or example dissection/analysis, to other psychological mechanisms, such as conceptual combination, or analogical abstraction; each of these mechanisms might then in turn be contextually important for certain kinds of creative problems, such as policymaking or room layout design. %Note that the elements in each coordinated space (e.g., contextualizing vs. \conPar viewing) appear to be linear and independent because of the design of our experiment; we expect that some higher-order patterns of relationships/pairings between patterns and/or mechanisms may be discovered.

We believe that fleshing out these paths through these coordinated spaces towards a theory of human-example interaction can both make contributions to fundamental HCI theory --- by enhancing synthesis of design knowledge about how to best support creative inspiration from examples --- and practice --- by providing a principled framework that is sufficiently granular and directly connected to design decisions, to guide effective design decisions when building example-based creativity support systems, and to practicing creators who wish to more effectively leverage examples in their creative process. %We are already finding this coordinated spaces framework useful for generating design hypotheses about effective example interaction designs that connect to other more sophisticated mechanisms like conceptual combination, which are at play in other sorts of creative task environments (e.g., policymaking, room layout design). 
We invite the rest of the creativity support systems community to join us in these efforts.

\bibliographystyle{ACM-Reference-Format}
\bibliography{sample-base}


\appendix
\newpage
\textbf{APPENDIX}

\section{Procedure for Generating Rugged WildCat Wells search environments}
\label{sec:appendixA}

We used four factors to control the synthetic objective functions: 
\begin{enumerate}
    \item The \textit{ruggedness amplitude}, in the range of [0,1], controlled the relative "height" of noise added to the search environment, compared to the height of the peaks. Increasing this parameter made the task more difficult by adding more places to incorrectly intuit as the location of the maximum reward. Setting this parameter to 1 would essentially make the search environment have infinite peaks. 
    \item The \textit{smoothness}, in the range of [0,1], controlled the degree of local correlation of scores in the grid. Intuitively, if smoothness was high (closer to 1), then a hill-climbing or gradient-based strategy could be viable, as a searcher that saw successive points in increasing score could (correctly) intuit that further points down that search path were likely to be of higher score; at lower values of smoothness, the search environment became very "bumpy", such that searchers would often be surprised by the score of nearby regions in the grid.
    \item The \textit{number of peaks} controlled the number of "maxima" in the search environment; intuitively, this specified the number of regions in the space where a searcher might (correctly or otherwise) intuit as the location of the treasure. Mathematically, this parameter controlled the number of layers of multivariate normal with single peaks.
    \item The \textit{distance between peaks}, in the range of [0,1], which prevented overlap of peaks when the function was generated with more than 1 peak.
\end{enumerate}

All search environments in this study were generated with ruggedness amplitude set to 0.7 (fairly noisy), smoothness level to 0.2 (fairly rugged), and the number of peaks to 1 with a maximum of 100; distance between peaks was not relevant because we only used a single peak, using the following algorithm:

\begin{algorithm}
\caption{Constructing the Wildcat Wells search environment with given ruggedness (noise) amplitude ($Rug_{amp}$), smoothness ($Smt$), number of peaks ($N$) and distance between peaks ($Rug_{freq}$).}
\begin{algorithmic}[1]
 \For {$Rug_{amp},Smt,N,Rug_{freq}$}\vspace{1ex}
 \State Get $X_{centers} = f(N,Rug_{freq})$
 \State Sample $\sum_i^N surf \sim \mathbb{N}(X_i)$ \vspace{1ex}
 \State Sample $Noise \sim \text{OpenSimplex}(Smt)$ \vspace{1ex}
 \EndFor \vspace{1ex}
 \State Return $surf+noise\times g(Rug_{amp})$.\vspace{1ex}
 \end{algorithmic}\label{alg:wildcatwells}
\end{algorithm}

\vspace{5ex}

% \section{Parameter Settings for Generating Rugged WildCat Wells Search Environments}
% \label{sec:appendixB}

\newpage
\section{Algorithm for Sampling Diverse and Non-Diverse Example Points}
\label{sec:appendixB}

\begin{algorithm}
\caption{Generating a ranked distribution of example sets with Determinantal Point Process (DPP) approach \cite{kulesza_determinantal_2012}, $M$ is batch size (10000). $S^k$ is a combinatorial set defined on a finite set $X \in \mathbb{R}^2$, where each element $S^k_{Y_i} \in S^k$ is k elements long.}
\begin{algorithmic}[1]
 \For {$i \in range(M)$}\vspace{1ex} 
 \State Sample $S^k_{Y_i} \sim \mathbb{IID}(S^k)$ [identically sampling unordered sets without replacement] \vspace{1ex}
 \State Calculate $g(S^k_{Y_i}) = g_{y_i}$ and append this to $Scores_{S^k}$ \vspace{1ex}
 \EndFor \vspace{1ex}
 \State Return DPP Score of sets of examples = $\frac{Score_{S^k} - mean(Score_{S^k})}{s.d.(score_{S^k})}$.\vspace{1ex}
 \end{algorithmic}\label{alg:dpp}
\end{algorithm}\vspace{0ex}

\end{document}